\documentclass[aps,apl,showpacs]{revtex4-1}
\usepackage{graphicx}

\usepackage{amsmath}
\usepackage{amsfonts}
\usepackage{latexsym}
\usepackage{natbib}
\usepackage{bm}
\usepackage{bbold}
\usepackage{color}

\newcommand{\I}{\mathrm{i}}
\newcommand{\e}{\mathrm{e}}
\begin{document}

\title{Sigma models for quantum chaotic dynamics}

\author{Alexander Altland}

\affiliation{Institut f{\"u}r Theoretische Physik,
Universit{\"a}t zu K{\"o}ln,  50937 K{\"o}ln, Deutschland}
\author{Sven Gnutzmann}
\affiliation{School of Mathematical Sciences, University of
  Nottingham, Nottingham, United Kingdom}
\author{Fritz Haake}
\affiliation{Fakult\"at f\"ur Physik, Universit\"at Duisburg-Essen,
  47048 Duisburg, Deutschland}

\author{Tobias Micklitz}
\affiliation{Centro Brasileiro de Pesquisas
    Fisicas, Rio de Janeiro, Brasil}

\begin{abstract}
  We review the construction of the supersymmetric sigma model for
  unitary maps, using the color-flavor transformation. We then
  illustrate applications by three case studies in quantum chaos. In
  two of these cases, general Floquet maps and quantum graphs, we show
  that universal spectral fluctuations arise provided the pertinent
  classical dynamics are fully chaotic (ergodic and with decay rates
  sufficiently gapped away from zero). In the third case, the kicked
  rotor, we show how the existence of arbitrarily long-lived modes of
  excitation (diffusion) precludes universal fluctuations and entails
  quantum localization.

\end{abstract}

\maketitle

\section{Introduction}
\label{intro}

We review the supersymmetric sigma model for unitary quantum maps,
including its derivation through the color-flavor transformation and present
three case studies of applications to quantum chaos. Our intention is
not to bring news to experts but rather to help newcomers getting
acquainted with and possibly using the method. Therefore,
we aim at a self-contained presentation.  Facing a choice, we treat
the simplest case and forgo generality. Following that motto
we confine our study to two-point quantities. The language of physics
rather than math will be spoken.

It has gradually transpired during the past decade that unitary maps
allow for easier and cleaner treatment than autonomous
flows. Zirnbauer's use of the color-flavor transformation to construct
a supersymmetric sigma model was an important first step
\cite{Zirnb96,Zirnb98,Zirnb99}. An advantage over previous treatments was
that the Hubbard-Stratonovich transformation as well as saddle-point
approximations could be avoided. Moreover, no regularization by
external noise was necessary. A step ahead was the in-depth treatment,
based on the sigma model, of the kicked rotor in all its ramifications
\cite{Tian10}. On the semiclassical side, the periodic-orbit treatment
of unitary maps could benefit from the existence of a rigorous variant
of the 'Riemann-Siegel lookalike' as well as of a finite trace formula
\cite{Braun12b}.

The sigma model is a field theoretic method expressing observables by
integrals over supermatrix valued 'fields'. It allows to look for the
sense in which and the conditions under which an individual quantum
system can display universal behavior, in its spectral, transport or
wave function characteristics. Even the study of system specific
properties is feasible. Of course, averages over ensembles are
possible as well.

A variant of the so called ballistic sigma model is accessible from
the general form. Employing a suitable
  Hilbert-space representation for the fields involved, like Wigner or
  Husimi functions, we can classically approximate the single-step
  quantum map. That approximation is acceptable provided the strobe
  period underlying the map is small compared to the Ehrenfest time
  since only on that latter scale qualitative discrepancies between
  quantum and classical dynamics arise. (In fact, most maps even have
  Lyapounov times larger than the strobe period.) The effectively
  classical single-step propagation does not impair the quantum nature
  of the Wigner or Husimi functions as quasi-probabilities.

The three case studies to be discussed refer to individual quantum
maps, rather than ensembles. For simplicity, they are all chosen
without time-reversal invariance. Quantum localization is at issue for
one of them (kicked rotor) while in the other two (general Floquet maps and
quantum graphs) we look at spectral fluctuations.

Before going to work we would like to spell out what can be expected
for spectral fluctuations of individual quantum systems.  The
two-point correlator of the density of levels is known not to be
self-averaging unless subjected to suitable smoothing. A recent
numerical investigation of smoothed versions of the two-point
correlator $R(e)$ for the kicked top \cite{Braun14} has revealed
self-averaging and fidelity to the pertinent RMT average (CUE)
provided the quasi-energy variable $e$ (reckoned in the mean level
spacing as a unit) remains within periodically repeated windows of
correlation decay. Within those windows a single spectrum makes for
but relatively tiny noise; outside, however, noise prevails and even
forbids distinction between CUE and GUE behavior.  Two smoothing
operations were shown to be equally useful in \cite{Braun14}. One is
integrating as $\int_0^e de'R(e')$ with Im\,$e=0^+$. The other is
complexification of the quasi-energy $e$ with ${1\over
  N}\ll\mathrm{Im}\,e\ll1$; the noise strength (rms) is $\sim
N^{-1/2}$ in the first case and $\sim (N\mathrm{Im}e)^{-1/2}$ in the
second.

The numerical results just alluded to do set limits to what we can
expect the sigma model to yield for individual dynamics. Most
importantly, we must introduce a smoothing operation before the sigma
model has a 'right' to a two-point correlator in any sense reminiscent
of the universal one. All the sigma model can and does produce for the
non-smoothed $R(e)$ of an individual system, is the correct train of
Dirac deltas engendered by the spectrum of the unitary (Floquet) map
$U$ (see Appendix \ref{app:existence}). Such singular behavior is
reflected in wild fluctuations of the sigma model fields. But
smoothing of the 'observable' $R(e)$ renders ineffective the wildness
and stabilizes the 'mean field' responsible for universal behavior,
given full chaos. On the other hand, the aforementioned $1\over\sqrt
N$ noise survives the smoothing, both in the numerical treatment of
Ref.~\cite{Braun14} and in the sigma-model representation of the
smoothed correlator $R(e)$. To capture such noise and its strength in
analytic terms we would have to scrutinize, as usual, the square of
the fluctuating quantity, going for a sigma-model based treatment of
the four-point function. Such an extension of our present work appears
quite feasable but will not be attempted here.

Our review is organized as follows. We first
(Sect.~\ref{sec:susymaps}) construct the sigma model by representing a
generating function $\mathcal Z$ (derivatives of which yield the
correlator $R(e)$) by an integral over supermatrix valued fields
$Z,\tilde Z$; the color-flavor transformation is the essential tool in
replacing a certain center-phase average required in the definition of
$\mathcal Z$ by the said superintegral. No approximation is involved
in the construction. --- Next (Sect.~\ref{sec:universal}), we give a
pedestrian evaluation of the generating function of the special case
of the 'zero dimensional' sigma model where the fields $Z$ and $\tilde
Z$ are stripped of any structure in the quantum Hilbert space
accommodating the Floquet matrix $U$, to become 'mean fields'. Those
mean fields afford the classical interpretation of ergodic equilibrium
in phase space. The resulting generating function is the one known
from the CUE of random-matrix theory. --- We proceed
(Sect.~\ref{sec:validityof0}) to revealing that the general sigma
model does indeed reduce to the zero dimensional one, provided the
classical limit of the quantum map is 'fully chaotic': its spectrum of
Ruelle-Pollicott resonances must exhibit a finite gap separating all
resonances from the single eigenvalue pertaining to the ergodic
equilibrium state; corrections turn out to be at most of the order
${1\over N}$; under certain conditions on the resonance spectrum the
corrections are even $\mathcal O({1\over N^2})$. At any rate, the
corrections are way smaller than the ${1\over\sqrt N}$ noise found in
Ref.~\cite{Braun14}. Our demonstration makes use of a strategy
proposed by Kravtsov and Mirlin \cite{Kravt94}. That strategy
separates the fields $Z$ and $\tilde Z$ into mean fields and
fluctuations in a way fully respecting the invariances of the
supermatrix manifold making up the sigma model. All of the work in
Sect.~\ref{sec:validityof0} requires protection by smoothing in the
sense mentioned above. --- In a short chapter (Sect.~\ref{sec:flows})
we argue that the sigma model for unitary maps can be used for
autonomous flows, simply by describing the latter stroboscopically
with a suitable strobe period. Some cleanliness is thus gained in
treating individual dynamics inasmuch as the aforementioned drawbacks
of previous 'ballistic sigma model' treatments are avoided. --- The
case study for the kicked rotor (Sect.~\ref{sec:rotor}) is meant to
show how the sigma model can cope with chaotic dynamics without a gap
in the resonance spectrum. We adhere to our motto of simplicity by
limiting ourselves to strong chaos (beyond the KAM regime), excluding
quantum resonances, and abstaining from including the well-known
renormalization of the diffusion constant. The remaining frame still
allows to show how arbitrarily long-lived diffusive modes make for
quantum localization.  --- Directed quantum graphs are treated in
Sect.~\ref{sec:graphs}. Here the restriction to incommensurate bond
lengths makes for a drastic simplification: the center-phase average
can be replaced by an $N$-fold average over the phases picked up by
the quantum wave along each of the $N$ bonds. Instead of the single
color-flavor transformation (one color, $N$ flavors) for general
Floquet maps, the graphs in question allow for $N$ such
transformations (one color, one flavor) such that the fields $Z$ and
$\tilde Z$ become block diagonal, both with a $2\times 2$ block for
each bond. Again separating into mean field and fluctuations and
treating the latter perturbatively we find the quantum dynamics taking
a classical appearance: The bond-to-bond scattering matrix appears
only with the squared moduli of its elements which afford
interpretation in terms of a classical Markov process. We are again
led to the validity of the mean-field treatment (equivalent to the
CUE) for large $N$ provided the spectrum of resonances is sufficiently
gapped. Corrections to the mean-field behavior then turn out to be at
most of the order $\propto{1\over N\Delta_g^2}$ with $\Delta_g$ the
said gap. As a special charm of the case, one gets the inkling that
rigorous treatments may be close at hand. --- The short section
\ref{sec:wrapup} wraps up what is achieved and what remains open.  

Several appendices provide supplementary material and aim at making
this review self-contained. In the first one (\ref{app:cft}), we
present a derivation of the color-flavor transformation as a more
pedestrian alternative to Zirnbauers beautiful original one
\cite{Zirnb96,Zirnb99}. The second appendix (\ref{app:flat}) explains
why the integration measure for our 'rational parametrization' in
terms of the fields $Z,\tilde Z$ is flat and the third
(\ref{app:Qinvariance}) expounds the origin of the general invariance
of the rational parametrization. The fourth appendix
(\ref{app:symspace}) puts the sigma model into the persective of group
theory and Riemann geometry. The fifth (\ref{app:existence}) is devoted
to easing existence worries; we here show that the correct train of
Dirac deltas arises for the non-smoothed correlator $R(e)$ when the
superintegral over the fields $Z,\tilde Z$ is done by picking up the
contributions of the so called standard and Andreev-Altshuler
saddles. Three further appendices fill in some calculations with which
we did not want to burden the main text, hoping to thus make for
better readability.

\section{Sigma model for Floquet maps, unitary symmetry}
\label{sec:susymaps}

{\bf Preliminaries:} We consider periodically driven quantum
systems. The time evolution of the wave function over one period of
the driving is effected by a unitary $N\times N$ 'Floquet' matrix $U$
without time reversal invariance nor any other symmetry.  Powers $U^n$
with $n=1,2,\ldots$ give a stroboscopic evolution. The unimodular
eigenvalues $\e^{-\I\varphi_\mu},\,\mu=1,2,\ldots, N$ are assumed, for
simplicity, to fill the interval $[0,2\pi]$ with the homogeneous mean
density $N/2\pi$. We define a 'microscopic' density
$\rho(\varphi)=\sum_\mu\delta_{2\pi}(\varphi-\varphi_\mu)$ with the
help of the $(2\pi)$-periodic delta function
$\delta_{2\pi}(\varphi)={1\over 2\pi}\sum_{n=-\infty}^\infty\e^{\I
  n\varphi}$. Fluctuations about the mean density are captured by the
two-point correlator
\begin{align}
  \label{eq:R(e)}
  R(e)=\left[\int_0^{2\pi}{d\phi\over 2\pi}\:
                  \rho\Big(\phi+{e\over N}\Big) \rho\Big(\phi-{e\over N}\Big)
                  -\Big({N\over 2\pi}\Big)^2\right]\Big({2\pi\over
                  N}\Big)^2
\,,
\end{align}
a normalized covariance averaged over the center phase $\phi$. The
function $R(e)$ is the real part of a 'complex correlator' $C(e)$
which can be written in terms of the eigenphases as
\begin{align}
  \label{eq:C(e)}
 C(e)={2\over N^2}\sum_{\mu,\nu}{\e^{\I (2e/N-\Delta_{\mu\nu})}
          \over 1-\e^{\I( 2e/N-\Delta_{\mu\nu})}}\,,\qquad\qquad 
\Delta_{\mu\nu}=\varphi_\mu-\varphi_\nu\,.
\end{align}
One is automatically led to $R=\mathrm{Re}\,C$ with the foregoing
$C(e)$ when expressing $R(e)$ in terms of the eigenphases of $U$.  The
limit Im\,$e\downarrow 0$ turns the real correlator
$R(e)=\mathrm{Re}\,C(e)$ into a train of ($2\pi$)-periodic Dirac
deltas, one for each eigenphase spacing $\Delta_{\mu\nu}$. Since the
$\Delta_{\mu\nu}$ have a mean distance ${2\pi\over N^2}$ an imaginary
part Im\,$e\gg {1\over N}$ broadens those delta functions sufficiently
to make the correlator structureless on the $e$-scale ${1\over N}$.

We shall eventually have to contrast the quantum 'Floquet map' $U$
with its classical counterpart, an area preserving map of the
classical phase-space density brought about by the so called
Perron-Frobenius operator $\mathcal F$. 'Iterations' $\mathcal F^n$
allow to follow how an initially smooth density approaches uniform
equilibrium as the dimensionless time $n$ grows, given ergodicity. The
equilibration looks most simple under coarse graining to limited
phase-space resolution. Exponential decay of smooth inhomogeneous
densities then arises with rates called Ruelle-Pollicott
resonances. It is also instructive to look at the classical map in the
'Newton picture' wherein a unique trajectory originates from any
initial phase-space point. Two trajectories starting at initially
close points separate, given chaos, exponentially with rates known as
Lyapounov exponents. The smallest Lyapounov rate visible in the Newton
picture corresponds to the smallest Ruelle-Pollicott resonance in the
'Liouville picture'.  Unfortunately, the definitions of the
correlators $R$ and $C$ do not lend themselves to direct use of the
chaotic character of the dynamics.

{\bf Generating function:} To get a chance to
  distinguish classically integrable and chaotic behavior we go
  towards the correlator through a generating function
\begin{equation}
  \label{eq:1}
  \mathcal Z(a,b,c,d)=\int_0^{2\pi}{d\phi\over 2\pi}\;\frac
{\det(1-c\e^{\I\phi}\,U)\det(1-d\e^{-\I\phi}\,U^\dagger)}
      {\det(1-a\e^{\I\phi}\,U) \det(1-b\e^{-\I\phi}\,U^\dagger)}
\end{equation}
which depends on four complex variables $a,b,c,d$; the moduli
$|a|,\,|b|$ are taken (infinitesimally) smaller than unity while $c,d$
remain unrestricted. The real 'center-phase' $\phi$ is averaged
over.

The foregoing definition entails the identities
\begin{align}
  \label{eq:23}
\mathcal Z(a,b,a,b)=1\qquad {\rm and}\qquad
\mathcal Z(a,b,c,d)= (cd)^{N}\mathcal Z\Big(a,b,{1\over d},{1\over c}\Big)
\end{align}
which we shall refer to as normalization and Weyl symmetry. Moreover,
the center-phase average has two consequences for any unitary operator
$U$, irrespective of symmetries and character of the limiting
classical dynamics. First, $\mathcal Z$ takes the value unity when
either $a=c$ for any $b,d$ or $b=d$ for any $a,c$ \cite{Braun14}, that is
\begin{align}
  \label{eq:Z-1}
  \mathcal Z-1\propto (a-c)(b-d)\,;
\end{align}
second, $\mathcal Z$ depends on $a,b,c,d$ only
through three combinations which can be chosen as
\begin{align}
  \label{eq:3combs}
  a=b=\e^{\I e/N}\,,\quad {c\over a}=\e^{\I\epsilon_+/N}\,,
  \quad \mathrm{and}\quad {d\over b}=\e^{\I \epsilon_-/N}
\end{align}
with $|ab|<1$ and ${c\over a},\,{d\over b}$ arbitrary complex.
We can therefore
extract the complex correlator from $\mathcal Z$ as
\begin{align}
  \label{eq:CfromZ}
  C(e)= \partial_c\partial_d \mathcal Z \,{2ab\over N^2}\Big|_{a=b=c=d=\e^{\I e/N}}
       =  -\partial_{\epsilon_+}\partial_{\epsilon_-} \mathcal Z\,
       {2ab}\Big|_{a=b=\e^{\I e/N},\epsilon_\pm=0}
={\mathcal Z-1\over (a-c)(b-d)}{2ab\over N^2}\Big|_{a=b=c=d=\e^{\I e/N}}\,.
\end{align}

The generating function carries physically relevant information only
for values of the variables $\epsilon_\pm$ close to zero, according to
(\ref{eq:CfromZ}). Higher-order derivatives
$\partial_{\epsilon_+}^m \partial_{\epsilon_-}^n\mathcal Z
\big|_{\epsilon_\pm=0}$ lead to higher-order correlators of the level
density as functions of the quasi-energy $e$ but also involve no more
than the $(\epsilon_\pm\approx 0)$-behavior. Moreover, nothing is lost
by restricting $\epsilon_\pm$ to real values. According to the
numerical results of Ref.~\cite{Braun14} the generating function
ceases to be faithful to RMT once $|\epsilon_\pm|\gtrsim 1$, so we
need not bother to look there.

Even the variable $e$ can eventually be taken as real, in the sense
Im\,$e\downarrow 0$. The latter limit brings forth delta function
singularities in $e$ for Re\,$\mathcal Z$ and principal-value type
${1\over e}$-singularities for Im\,$\mathcal Z$. As already pointed
out in the introduction, however, keeping an imaginary part of $e$ as ${1\over
  N}\ll\mathrm{Im}e\ll1$ is one of the possible smoothings endowing the
correlator $C(e)$ with the property of self-averaging and providing
protection for the perturbative treatment of the sigma model we shall
have to undertake.

It seems we are not any closer yet to distinguishing regular and
chaotic motion. A reformulation of the definition of the generating
function --- to what is called the sigma model --- will help.

{\bf Sigma model:} To proceed towards the sigma model we represent the
ratios of determinants in (\ref{eq:1}) by Gaussian integrals and write
the generating function as
\begin{eqnarray}
  \label{eq:2}
  \mathcal Z = \int_0^{2\pi}{d\phi\over 2\pi}\;\int d(\psi,\psi^*)
\;\exp\sum_{k,l=1}^N
\Big\{-\psi_{+,B,k}^*(\delta_{kl}-a\e^{\I\phi}\,U_{kl})\psi_{+,B,l}
             -\psi_{-,B,k}^*(\delta_{kl}-b\e^{-\I\phi}\,U_{lk}^*)\psi_{-,B,l}&
\nonumber\\
             -\,\psi_{+,F,k}^*\,(\delta_{kl}-c\e^{\I\phi}\,U_{kl})\psi_{+,F,l}
             -\psi_{-,F,k}^*\,(\delta_{kl}-d\e^{-\I\phi}\,U_{lk}^*)\psi_{-,F,l}&
\;\Big\}\,;
\end{eqnarray}
here, the $4N$ integration variables comprise $2N$ pairs of mutually
complex conjugate ordinary (Bosonic) variables
$\psi_{\pm,B,k},\psi_{\pm,B,k}^*$ and $4N$ mutually independent
Grassmannians $\psi_{\pm,F,k},\psi_{\pm,F,k}^*$. The first index,
$\lambda=+,-$, distinguishes the two 'columns' in the definition
(\ref{eq:1}) of $\mathcal Z$ which can be associated with forward
($+$) and backward $(-)$ time evolution; we shall speak of the two
dimensional advanced ($-$)/retarded (+) space AR when referring to that
index. The second index, $s={B,F}$, distinguishes denominator and
numerator in (\ref{eq:1}); since the two cases are respectively
represented by Bosonic and Fermionic variables in the integral
(\ref{eq:2}) we speak of a two dimensional Bose-Fermi (BF) space.
Finally, the third index, $k=1,2,\ldots N$, pertains to the Hilbert
space of the quantum dynamics (QD) wherein the Floquet matrix
operates. The flat integration measure reads
\begin{equation}
\label{eq:3}
d(\psi,\psi^*)=\prod_{k=1}^N\frac{d^2\psi_{+,B,k}}{\pi}\frac{d^2\psi_{-,B,k}}{\pi}\,
d\psi_{+,F,k}\,d\psi_{+,F,k}^*\,d\psi_{-,F,k}d\,\psi_{-,F,k}^*\,.
\end{equation}
The mentioned restrictions on $a,b$ secure the existence of the
Bosonic integrals in (\ref{eq:2}).  We may lump the integration
variables into four supervectors $\psi_\pm$ and $\psi_\pm^*$ and
introduce the diagonal BF matrices
\begin{align}
  \label{eq:epm}
  \hat e_+=\Big({a\atop }\,{\atop c}\Big)
   \,,\qquad
  \hat e_-=\Big({b\atop }\,{\atop d}\Big) 
\end{align}
and compact the superintegral (\ref{eq:2}) to
\begin{eqnarray}\label{eq:4}
\mathcal Z=\int_0^{2\pi}{d\phi\over 2\pi}\;\int d(\psi,\psi^*)\;\exp
\Big\{
-\psi_+^{*T}\big(1-\e^{\I\phi}\hat e_+U\big)\psi_+
-\psi_-^{*T}\big(1-\e^{-\I\phi}\hat e_-U^\dagger\big)\psi_-
\Big\}
\,.
\end{eqnarray}

The center-phase average
$\langle\cdot\rangle=\int_0^{2\pi}\frac{d\phi}{2\pi}(\cdot)$ can be
traded against a supermatrix average, using the 'color-flavor
transformation' \cite{Zirnb96,Zirnb98,Zirnb99} (for a derivation, see
App.~\ref{app:cft}). For arbitrary supervectors
$\Psi_1,\Psi_2,\Psi_{1'},\Psi_{2'}$ that transform reads
\begin{equation}
  \label{eq:cft}
\int_0^{2\pi}\frac{d\phi}{2\pi} \;\exp
\big\{\e^{\I\phi}\Psi_1^T\Psi_{2'}+\e^{-\I\phi}\Psi_2^T\Psi_{1'}\big\}
= \int \!d(Z,\tilde Z) \;\mathrm{sdet}(1-Z\tilde Z)\,
\exp\big\{\Psi_1^T\tilde Z\Psi_{1'}+\Psi_2^T Z\Psi_{2'}\big\}\,.
\end{equation}
The integration measure $d(Z,\tilde Z)$ is flat (see Appendix
\ref{app:flat}), defined analogously to (\ref{eq:3}), and
$\mathrm{sdet}$ denotes the superdeterminant (see
e.g. \cite{Haake10}). The Bose-Bose and Fermi-Fermi blocks of the
matrices $Z, \tilde Z$ are subject to the (convergence generating)
conditions ${\tilde Z}_{BB}=Z_{BB}^\dagger,\,{\tilde
  Z}_{FF}=-Z_{FF}^\dagger$ and the integration domain is restricted as
$|Z_{BB}Z_{BB}^\dagger|<1$, that is, all eigenvalues of
$Z_{BB}Z_{BB}^\dagger$ lie in the interval $[0,1)$.

For our study of the unitary symmetry class the supervectors in the
integral transformations (\ref{eq:cft}) have $2N$
components. Correspondingly, the supermatrices $Z,\tilde Z$ are
$2N\times 2N$ and act in BF$\otimes$QD. The transformation
(\ref{eq:cft}) entails
\begin{eqnarray}
\label{eq:7}
  \mathcal Z&=& \int d(\psi,\psi^*)\;
\exp\big(-\psi_+^{*T}\psi_+-\psi_-^{*T}\psi_-\big)
\int d(Z,\tilde Z) \;\mathrm{sdet}(1-Z\tilde Z)\,
\exp\big(\psi_+^{*T}\tilde{ Z}\psi_- +
\psi_-^{*T}U^\dagger\hat e_- Z\hat e_+U\psi_+\big)\,.
\end{eqnarray}
To compact even further we introduce the AR$\otimes$BF$\otimes$QD
vectors $\psi=\big({\psi_+\atop \psi_-}\big),\,\psi^*=\big({\psi_+^*\atop
  \psi_-^*}\big)$ and the matrix
\begin{equation}
  \label{eq:9}
  M= 1-\Big({0\atop \hat e_- Z\hat e_+}\,{U\tilde Z U^\dagger\atop 0}\Big)_{\!\mathrm{AR}}\,.
\end{equation}
Now, the Gaussian integral over $\psi$ and $\psi^*$  can be
done,
\begin{equation}
\label{eq:8}
\int d(\psi,\psi^*)\,\exp(-\psi^\dagger M\psi)
=\big(\mathrm{Sdet} \, M\big)^{-1}
=\exp\big(-\mathrm{Str}\ln M\big)\,,
\end{equation}
where Str and Sdet refer to the $4N$ dimensional space
AR$\otimes$BF$\otimes$QD (we reserve sdet and str to BF$\otimes$QD).
Expanding $\ln M$ in powers of $M-1$, observing that the supertrace of
odd powers of $M-1$ vanishes and resumming one gets $\mathrm{Str}\ln M=
\mathrm{str}\ln (1-\tilde Z \hat e_-U^\dagger Z \hat e_+U)$. We have
arrived at the \textit{sigma model} for unitary quantum maps from the
unitary symmetry class,
\begin{align}
  \label{eq:10}
  \mathcal Z=&\int d(Z,\tilde Z)\e^{- \mathcal S(Z,\tilde Z)}\\
  \label{eq:11}
  \mathcal S(Z,\tilde Z)=&-\mathrm{str}\ln\big(1-\tilde Z Z\big)
+\mathrm{str}\ln\big(1-U\tilde Z U^\dagger \hat e_-Z\hat e_+\big)\,;
\end{align}
the action $\mathcal S$ is the central object of the
theory. The measure is normalized as
$\int d(Z,\tilde Z)\e^{\mathrm{str}\ln\big(1-\tilde Z Z\big)}=1$. We
are obviously free to change names as $Z\leftrightarrow\tilde Z$.

At this point we can see that the generating function depends only on
three independent variables which we may choose as in
(\ref{eq:3combs}). To reveal that property in the action $S(Z,\tilde
Z)$, we write the $2N\times 2N$ supermatrix $\hat e_-Z\hat e_+$ as
$2\times 2$ in BF (with the entries $N\times N$ in QD) and find
\begin{align}
  \label{eq:38}
\hat e_-Z\hat e_+=
           \left({abZ_{BB}\atop adZ_{FB}}{bcZ_{BF}\atop cdZ_{FF}}\right)=
           ab\left({Z_{BB}\atop {d\over b}Z_{FB}}\,
           {{c\over a}Z_{BF}\atop {c\over a}{d\over b}Z_{FF}}\right)\,.
\end{align}

The action can be rewritten as a supertrace in
AR$\otimes$BF$\otimes$QD. To do that
we take the matrix $M$ of
(\ref{eq:9}) as $2\times 2$ in AR  with $(2\times 2)\otimes
(N\times N)$ blocks in  BF$\otimes$QD.  Some reshuffling (see
App.~\ref{sec:susySwithQ}) gives
\begin{align}\label{eq:susySwithQ}
\mathcal S(Z,\tilde Z)&=
\mathrm{Str}\ln \left[1+\Lambda (1-\hat U)(1+\hat U)^{-1}Q(Z,\tilde Z)\right]
\,{+ \;\mathrm{Str}\ln (1+\hat U)}
,\\ \nonumber
\\\label{eq:Q}
Q(Z,\tilde Z)&=\left({1\atop\tilde  Z}{ Z\atop 1}\right)
                    \left({1\atop 0}{0\atop -1}\right)
                    \left({1\atop\tilde  Z}{ Z\atop 1}\right)^{-1}
\equiv T\Lambda T^{-1}\,,\qquad
\hat U=\left({U \hat e_+\atop }{\atop U^\dagger \hat e_- }\right)\,.
\end{align}
We note that the matrix $Q$ arising from the 'trivial' configuration
$\Lambda$ by the transformation $T$ obeys $\mathrm{Str}\,Q=0$ and
$Q^2=1$. The manifold explored by $Q$ as $Z$ and $\tilde Z$ vary is
a symmetric space whose properties are discussed in the Appendices
\ref{app:flat}, \ref{app:Qinvariance}, and \ref{app:symspace}.

The sigma model, be it in the form (\ref{eq:10},\ref{eq:11}) or
(\ref{eq:susySwithQ},\ref{eq:Q}), does finally allow us to make
explicit use of the chaos assumed for the classical dynamics, simply
because the action $\mathcal S$ involves the Floquet matrix $U$ only
through the single-step time evolution $U\tilde Z U^\dagger$. Assuming
the underlying period of the driving smaller than the smallest
Lyapounov (or Ruelle-Pollicott) time we can classically approximate
the single-step quantum evolution, quantum corrections vanishing like
a power of $\hbar$ in the semiclassical limit $\hbar\sim{1\over N}\to
0$. Only on the much larger Ehrenfest time scale would quantum
features of the dynamics become dominant. With the said approximation
we shall arrive at the \textit{ballistic sigma model}.
 
The original representation (\ref{eq:11}) of the action is the natural
starting point for studying the universal limit. It will also serve us
fine for our investigation of deviations from universality; however,
the alternative (\ref{eq:susySwithQ},\ref{eq:Q}) will turn out
indispensable for identifying the parts of $Z,\tilde Z$ which can be
the basis of a systematic treatment of deviations.

We conclude our introduction of the sigma model by pointing out that
not only the generating function $\mathcal Z$ but also the correlator
$C(e)$ affords a representation by a superintegral. To see this we
expand the action $\mathcal S$ as $\mathcal S = \mathcal
S_0+(a-c)\mathcal S_++(b-d) \mathcal S_- +(a-c)(b-d)\mathcal
S_{+-}+\ldots$ and immediately proceed to the ensuing expansion for the
generating function, $\mathcal Z=\int d(Z,\tilde Z)\,\e^{-\mathcal
  S_0}\,\big[1- (a-c) S_+-(b-d) \mathcal S_- +(a-c)(b-d)(\mathcal
S_+\mathcal S_- -\mathcal S_{+-})\big]+\ldots$.  Now due to the identity
(\ref{eq:Z-1}) the zero-order term ('1' in the forgoing square bracket)
integrates to unity while the first-oder terms integrate to
zero. Finally, using the last member in (\ref{eq:CfromZ}) we arrive
at the promised sigma-model representation of the correlator,
\begin{align}
  \label{eq:18}
C(e)= {2\e^{\I e/N}\over N^2}\int d(Z,\tilde Z)
         \,\e^{-\mathcal S_0}\, \big(\mathcal S_+\mathcal S_+-S_{+-}\big)\,.
\end{align}
The work to be presented could be simplified by right away forgetting
about the generating function. However, we prefer to hold on to
$\mathcal Z$ for a while, for the simple reason already mentioned
above: there is physics in the higher-order derivatives
$\partial_{\epsilon_+}^m \partial_{\epsilon_-}^n\mathcal
Z\big|_{\epsilon_\pm=0}$.  Only near the end of our various
investigations of $\mathcal Z$ we shall resign to discussing no more
than the two-point correlator and then benefit from the pertinent
simplification.

\section{The universal limit: zero dimensional sigma model}
\label{sec:universal}

As a first application, we reduce the sigma model to the so called
zero dimensional one and present an elementary evaluation of its
generating function. We shall recover the universal spectral
fluctuations as encapsulated in the Circular Unitary Ensemble of
random-matrix theory \cite{Mehta04}.

\subsection{Formal reduction}
\label{subsec:formalred}

The reduction in question requires that the matrices $Z,\tilde Z$ act
like unity in the QD Hilbert space. Given
\begin{equation}
  \label{eq:ZtoB}
  Z\to \mathbb{1}_{\mathrm{QD}} \,B\,, \qquad \tilde Z\to
  \mathbb{1}_{\mathrm{QD}} \,\tilde B
\end{equation}
with $2\times 2$ supermatrices $B,\,\tilde B$, the Floquet operator
disappears due to $UU^\dagger=1$. This is how system
independent universal behavior is invited. The supertrace in
(\ref{eq:11}), over both  BF and QD, then just yields a
factor $N$ from QD and only the BF supertrace remains.  The
generating function becomes an integral over the $2\times 2$
supermatrices $B,\tilde B$
\begin{equation}
  \label{eq:13}
  \mathcal Z_0=
    \int d(B,\tilde B)\;\e^{-N\mathcal S_0(B,\tilde B)}\,, \qquad
    \mathcal S_0(B,\tilde B)=-\mathrm{str}\ln (1-\tilde B B)
    +\mathrm{str}\ln (1- \tilde B\hat e_-B\hat e_+)\,,
\end{equation}
with the supertrace from this point on refering only to the two
dimensional BF space.  The subscript '0' on the generating function
signals a certain zero dimensional character of the model due to the
absence of Hilbert space structure; we shall drop that subscript in
the remainder of the present section. At any rate, we have arrived at
the so called zero dimensional sigma model for the circular unitary
ensemble (CUE) of random matrices.  The resulting explicit form of the
generating function at finite $N$ was rigorously determined as a CUE
average in \cite{Conre07} and semiclassically constructed for
individual dynamics in the framework of periodic-orbit theory
\cite{Braun12b}.

\subsection{Generating function}
\label{subsec:polar}

One may parametrize the
supermatrices $B$ and $\tilde B$ by their matrix elements,
\begin{align}
  \label{eq:24}
  B=\Big({s\atop \nu}{\mu \atop t} \Big)\,,\qquad\tilde B=\Big({s^*\atop
  \mu^*}{\nu^* \atop -t^*}\Big)\,.
\end{align}
The integration ranges then  are the unit disc for $s$ and the whole complex
$t$-plane, and the integration measure is flat.
A more convenient set of integration variables is provided by the
singular-value decomposition
\begin{align}
  \label{eq:singvaldec}
  B=W
  \begin{pmatrix}
   \sqrt l_B&\\ &\sqrt{- l_F}
  \end{pmatrix}V^{-1}\equiv W\underline{B} V^{-1}
\,, \qquad
 \tilde B=V
  \begin{pmatrix}
    \sqrt l_B&\\ &-\sqrt{-l_F}
  \end{pmatrix}W^{-1}\equiv V\underline{\tilde B}W^{-1}\,;
\end{align}
here $l_B$ and $-l_F$ both have positive numeric parts (see
\cite{Haake10}, p.535); the roots $\sqrt l_B$ and $\sqrt{- l_F}$ are
meant positive. It follows that the products $B\tilde B$ and $\tilde B
B$ are diagonalized as $W^{-1}B\tilde B W=V^{-1}\tilde B
BV=\big({l_B\atop }{\atop l_F}\big)$. The supermatrices $W$ and $V$
are given by 
\begin{align}
  \label{eq:59}
  W&=
  \left( {1+\frac{\eta\eta^*}{2}\atop\eta^*}
        {\eta \atop1-\frac{\eta\eta^*}{2}}\right)\left({\e^{-\I\varphi_B\atop}}{\atop
        \e^{-\I\varphi_F}}\right)\,,\qquad
  V=
  \left( {1-\frac{\tau\tau^*}{2}\atop-\tau^*}
        {\tau\atop1+\frac{\tau\tau^*}{2}}\right)\,,\\
 W^{-1}&=\left({\e^{\I\varphi_B\atop}}{\atop
        \e^{\I\varphi_F}}\right)
  \left( {1+\frac{\eta\eta^*}{2}\atop -\eta^*}
        {-\eta \atop1-\frac{\eta\eta^*}{2}}\right)\,,\qquad\,
  V^{-1}=
  \left( {1-\frac{\tau\tau^*}{2}\atop\tau^*}
        {-\tau\atop1+\frac{\tau\tau^*}{2}}\right)\,.
\end{align}
The singular values $ \sqrt l_B,\sqrt{- l_F}$, the two phases
$\varphi_B,\varphi_F$, and the four independent Grassmannians
$\eta,\eta^*,\tau,\tau^*$ can be expressed in terms of the matrix
elements of $B,\tilde B$ \cite{Haake10}. Taking these eight new parameters
(sometimes called 'polar coordinates') as integration variables we get
the integration measure and the integration range \cite{Haake10}
\begin{align}
  \label{eq:65}
  d(B,\tilde B)=\frac{dl_Bdl_F d\varphi_B d\varphi_F d\eta^*d\eta d\tau^* d\tau}{4\pi^2(l_B-l_F)^2}
\,,\qquad 0< l_B\leq 1\,,\quad -\infty<l_F< 0\,,\quad 0\leq
\varphi_B,\varphi_F\leq 2\pi\,.
\end{align}

Using the decomposition (\ref{eq:singvaldec}) we can write the two terms of the action in
(\ref{eq:13}) as
\begin{align}
  \label{eq:60}
 \mathcal S_1=& -\mathrm{str }\ln (1-\tilde B B)=\ln\frac{1-l_F}{1-l_B}\,,\\
  \mathcal S_2=&\;\mathrm{str}\ln (1- \tilde B\hat e_-B\hat e_+)=
 \mathrm{str}\ln \left(1-\underline{\tilde B}W^{-1}\hat e_-W\underline B
V^{-1}\hat e_+V\right)\,.
\end{align}
To process the second term, we note
\begin{align}
  \label{eq:61}
  V^{-1}\hat e_+V&=\hat e_+ + (a-c)
  \left({-\tau\tau^*\atop \tau^*}{\tau\atop -\tau\tau^*}\right)
\equiv \hat e_++(a-c)\Delta_+(\tau,\tau^*)
\\ \nonumber
 W^{-1}\hat e_-W&=\hat e_- + (b-d)
  \left({\eta\eta^*\atop -e^*\eta^*}\;{e\eta\atop \eta\eta^*}\right)
\equiv \hat e_- +(b-d)\Delta_-(\eta,\eta^*)\,;
\end{align}
nilpotent matrices $\Delta_{\pm}$ occur here, and $\Delta_-$ contains
the unimodular parameters $e\equiv\e^{\I(\varphi_B-\varphi_F)}$. These
identities yield $\mathcal S_2=\mathrm{str }\ln \left(1-
  \underline{\tilde B}\big(\hat e_-+(b-d)\Delta_- \big) \underline{B}
  \big(\hat e_++(a-c)\Delta_+\big)\right)$.  Extracting the diagonal
matrix $\left({1-abl_B\atop }{ \atop 1-cdl_F}\right)$ as a factor from
the argument of the foregoing logarithm we split $\mathcal S_{2}$ into
a numerical and a nilpotent summand as
\begin{align}
  \label{eq:64}
  \mathcal S_2&=\ln\frac{1-abl_B}{1-cdl_F}+\mathrm{str }\ln (1-m)\,,
\\ \label{eq:m}
     m(\tau,\tau^*,\eta,\eta^*)&=\textstyle{\Big({(1-abl_B)^{-1}\atop}
               {\atop (1-cdl_F)^{-1}}\Big)}
                    \big( (a-c)\underline{\tilde
                       B}\,\underline{B}\hat e_-\Delta_+
                   +(b-d) \underline{\tilde B}\Delta_-
                  \underline{B}\hat e_+
                   +(a-c)(b-d) \underline{\tilde B}\Delta_-
                    \underline{B} \Delta_+
            \big)
\end{align}
We shall have to look more closely at the nilpotent matrix $m$ presently.

We can dispose of the unimodular variables $e,e^*$; according to
(\ref{eq:61}) they appear only in the offdiagonal elements of the
matrix $\Delta_-$ and disappear when we change integration variables
as $e\eta\to \eta, e^*\eta^*\to\eta^*$.  The integrals over the phases
$\phi_B$ and $\phi_F$ then just give the factor $4\pi^2$.  Collecting
everything we write the generating function
\begin{align}
  \label{eq:66}
  \mathcal Z=&\lim_{\epsilon\to 0} 
\int_\epsilon^1  dl_B\int_{-\epsilon}^{-\infty} dl_F
  \left[\frac{1}{(l_B-l_F)^2}
 \left(\frac{(1-l_B)(1-cdl_F)}{(1-l_F)(1-abl_B)}\right)^N
 \int d\eta^*d\eta d\tau^* d\tau\,\e^{-N\mathrm{str}\,\ln(1-m)}\right]
\end{align}
where from here on the matrix $\Delta_-$ entering $m$ is understood
purged of the said phase factors, i.~e.~, $e\to 1$.

We need to explain why we have cut an $\epsilon$-neighborhood of
$l_B=l_F=0$ from the integration range and defined the integral
through the limit $\epsilon\to 0$. According to the transformation
(\ref{eq:singvaldec}), the 'point' $l_B=l_F=0$ corresponds to
$B=\tilde B=0$ where the integrand has no singularity in terms of the
parametrization (\ref{eq:24}), such that cutting an infinitly small
neighborhood does not alter the integral. Moreover, the integral
exists independently of the chosen integration variables and must
therefore be correctly given by the limiting procedure in
(\ref{eq:66}). On the other hand, the transformation (\ref{eq:singvaldec})
does bring in a singularity at $l_B=l_F=0$.  Indeed, the matrix $m$
vanishes there while the factor $\frac{1}{(l_B-l_F)^2}$ diverges. As a
consequence, the integral $\int_0^1dl_B\int_0^{-\infty}dl_F[\ldots]$
differs from $\mathcal Z=\lim_{\epsilon\to 0} \int_\epsilon^1
dl_B\int_{-\epsilon}^{-\infty} dl_F[\ldots]$ by a finite term due to
the singularity of the integrand in the 'polar coordinates'. We shall
not have to labor to get the contribution of that singularity since it
is automatically produced by the general property (\ref{eq:Z-1}) as
unity.  We conclude
\begin{align}
\label{eq:intZ-1}
   \mathcal Z-1=\int_0^1dl_B\int_0^{-\infty}dl_F
     \left[\frac{1}{(l_B-l_F)^2}
 \left(\frac{(1-l_B)(1-cdl_F)}{(1-l_F)(1-abl_B)}\right)^N
 \int d\eta^*d\eta d\tau^* d\tau\,\e^{-N\mathrm{str}\,\ln(1-m)}\right]\,.
\end{align}
We can even conclude that the factor $(a-c)(b-d)$ in $\mathcal Z-1$ must
come from the Grassmann part of the foregoing superintegral. An
elementary if somewhat lengthy calculation (see App.~\ref{app:intZ0})
gives that Grassmann integral as
\begin{align}
  \label{eq:G}
  \mathcal G= \int d\eta^*d\eta d\tau^*d\tau\,\e^{-N\mathrm{str}\,\ln(1-m)}
                  = (a-c)(b-d) {N(l_B-l_F)\big(1-abcdl_Bl_F+N(abl_B-cdl_F)\big)
                        \over (1-abl_B)^2(1-cdl_F)^2}\,.
\end{align}
The generating function $\mathcal Z$ can then be found without doing
the remaining twofold integral over $l_B$ and $l_F$, by just invoking
the Weyl symmetry. We rewrite that symmetry for the quantity
\begin{align}
  \label{eq:136}
  J(ab,cd)= \frac{(1-ab)(1-cd)}{(a-c)(b-d)}(\mathcal Z-1)
\end{align}
which depends only on the two variables $ab$ and $cd$, and confront
the identity
\begin{align}
  \label{eq:145}
 \frac{(a-c)(b-d)}{(1-ab)(1-cd)}\Big(J(ab,cd)+(cd)^N J(ab,\textstyle{1\over cd})\Big)
 =(cd)^N-1-(cd)^N J(ab, \textstyle{{1\over cd}})\,.
\end{align}
Herein, the r.~h.~s.~deppends only on the variables $ab,cd$ while the
l.~h.~s.~is proportional to the differences $a-c$ and $b-d$. So
both sides vanish. 
We conclude $J=1-(cd)^N$ and arrive at the known generating function
of the CUE
\begin{align}
  \label{eq:0DZ}
{\mathcal Z} =1+\Big(1 -(c d)^N\Big) { (a-c) (b - d) \over (1 - a b)(1 -  c d)}
                   ={\mathcal Z}_{\mathrm{CUE}}\,.
\end{align}

\subsection{Directly to the correlator}
\label{sub:corrdirect}
A lot of work can be saved by directly going to the correlator
$C(e)$.  According to the last member of Eq.~(\ref{eq:CfromZ}) we can
set $a=b=c=d$ in the cofactor of $(a-c)(b-d)$ in the Grassmann
integral $\mathcal G$ given in (\ref{eq:G}).  To do the thus
simplified Grassmann integral we simply decompose the matrix $m$ as
$m=(a-c) m_++(b-d) m_-+(a-c) (b-d) m_{+-}$, reading the
summands from the definition (\ref{eq:m}). We then simplify the logarithm
$\ln(1-m)$ by keeping only terms capable of contributing a term
$\propto (a-c)(b-d)$ to $\mathcal G$. We thus get
$\e^{-N\mathrm{str}\,\ln(1-m)}\to N (a-c)(b-d)
\big\{\mathrm{str}\,(m_{+-}+m_+m_-) +
N\mathrm{str}\,m_+\mathrm{str}\,m_-\big)\}$\,.  The subsequent Grassmann
integration is survived only by terms $\propto \eta\eta^*\tau\tau^*$
within the foregoing curly bracket; that fact is easily checked to
allow for dropping the skew entries in $\Delta_\pm$. We thus
immediately get
\begin{align}
  \label{eq:154}
  \mathcal G \to (a-c)(b-d){N(l_B-l_F)\over (1-a^2l_B)^2 (1-a^2l_F)^2}
  \big\{ 1-a^4l_Bl_F+Na^2(l_B-l_F)\big\}
\end{align}
and the correlator
\begin{align}
  \label{eq:155}
  C(e)=&\,{2a^2\over N}\int_0^1dl_B\int_{-\infty}^0dl_F\;
          \Big({1-l_B\over 1-l_F}\Big)^N{(1-a^2l_F)^{N-2}\over (1-a^2l_B)^{N+2}}
          \; {1-a^4l_Bl_F+Na^2(l_B-l_F)\over l_B-l_F}
          \equiv {2a^2\over N}(1+a^2\partial_{a^2})\,h(a^2)\\ \nonumber
\\ \nonumber
h(z)=&\,\int_0^1dl_B\int_{-\infty}^0dl_F\;
          \Big({1-l_B\over 1-l_F}\Big)^N\;{(1-zl_F)^{N-1}\over (1-zl_B)^{N+1}}
          \; {1\over l_B-l_F}\,,
\end{align}
where $z=a^2=\e^{\I2 e}$. The remaining twofold integration cannot be
cut short by the Weyl symmetry but is easily done as follows. We
expand the integrand in powers of $z$ to obtain
$h(z)=\sum_{n_2=0}^{N-1}\sum_{n_1=0}^\infty f_{n_1n_2} z^{n_1+n_2}$
and $(1+z\partial z)h(z)=\sum_{n_2=0}^{N-1}\sum_{n_1=0}^\infty
(1+n_1+n_2)f_{n_1n_2} z^{n_1+n_2}$ with
\begin{align}
  \label{eq:5}
  (1+n_1+n_2)f_{n_1n_2}=(1+n_1+n_2)\left(N-1\atop n_2\right) \left(N+n_1\atop n_1\right)
              \int_0^1dl_B\int_{-\infty}^0dl_F \Big({1-l_B\over 1-l_F}\Big)^N
              {(-l_F)^{n_2}l_B^{n_1}\over l_B-l_F}=1\,.
\end{align}
To check that the latter coefficients all equal unity we change
integration variables, first as $l_F\to x={1-l_B\over 1-l_F}$ and
subsequently as $l_B\to (1-x)y$. Those transformations bring the
foregoing double integral over $l_B,l_F$ to the form of a product of
two separate integrals like $\int_0^1x^n(1-x)^m={m!n!\over
  (n+m+1)!}$\,, and the result $(1+n_1+n_2)f_{n_1n_2}=1$ arises. The
correlator thus becomes the product of two geometric series,
$C={2z\over N}\sum_{n_2=0}^{N-1}\sum_{n_1=0}^\infty z^{n_1+n_2}$,
producing the final form
\begin{align}
  \label{eq:CCUE}
  C(e)={e^{2\I e}-1\over 2N^2\sin^2{e\over N}}
       =C_{\mathrm{CUE}}(e)\,.
\end{align}

We face periodicity in the phase $e$ with period $\pi N$. Much smaller
is the phase scale in which $C(e)$ takes values independent of $N$,
according to $N^2\sin^2{e\over N}\sim e^2$ and $C(e)\sim {e^{2\I
    e}-1\over 2e^2}$, in coincidence with the correlator of the
Gaussian unitary ensemble. That behavior arises in windows of
'correlation decay and revival', around
$e=0,N\pi,2N\pi,\ldots$. Outside those windows $\sin^2{e\over N}$
defies Taylor expansion and the correlator is of the order ${1\over
  N^2}$.

\section{Limit of validity of the zero dimensional model}
\label{sec:validityof0}

\subsection{Preparatory remarks}
\label{subsec:prepremarks}

We have seen the zero dimensional sigma model to discard all QD
structure of the matrices $\tilde Z, Z$, cf.~(\ref{eq:ZtoB}). Now, we
want to complement the 'mean field' $\mathbb{1}_{\mathrm{QD}} \,B$ by
fluctuations. A quantitative treatment of that complement will reveal
conditions under which deviations from universal spectral fluctuations
are negligible.

A clue will be provided by a representation for $\tilde Z, Z$ allowing
to implement the semiclassical limit, such that the single-step
quantum evolution $U\tilde ZU^\dagger$ can be classically
approximated. The error thus incurred is small provided the strobe
period underlying the single-step Floquet map is small against the
Ehrenfest time $t_E$, the time scale on which the quantum evolution
acquires qualitative discrepancies from classical behavior.

The condition for universality of spectral fluctuations of a single
quantum system will turn out to be that there must not be any
arbirarily long-lived mode in the classical limit. Rather, all
classical Frobenius-Perron rates must be finitely gapped away from the
eigenvalue zero. The latter pertains to the unique ergodic equilibrium
with uniform probability density covering (the accessible part of) the
classical phase space. That ergodic equilibrium is the classical
counterpart of the absence of any QD structure in the configurations
(\ref{eq:ZtoB}) admitted by the zero dimensional sigma model. The
absence of any conservation laws both for the quantum and the classical
dynamics is implicitly assumed and in fact part of the conditions for
universal fluctations in the quantum spectrum.  The conditions for
spectral universality just anticipated have also be found as
constitutive for the periodic-orbit based theory of spectral
fluctuations \cite{Muell09,Braun12b}.

The following \textit{calculations must be done
  under the protection of sufficient smoothing}, or else the
correlator and the generating function do not even exist as ordinary
functions. Complexifying the quasienergy  $e$ suffices \cite{Braun14}.

\subsection{Perturbatively tractable fluctuating fields}
\label{subsec:fluctuations}
For an intuitive generalization of the reduction (\ref{eq:ZtoB}) we
split the matrices $\tilde Z, Z$ into mean fields and fluctuations as
\begin{align}
  \label{eq:12}
  Z=\mathbb{1}_{\mathrm{QD}} \,B+\delta Z\,,\qquad
  \tilde  Z=\mathbb{1}_{\mathrm{QD}} \,\tilde B+\delta \tilde Z\,.
\end{align}
An alternative possibility pioneered by Kravtsov and Mirlin
\cite{Kravt94} is offered by the form
(\ref{eq:susySwithQ},\ref{eq:Q}) of the action $S$: we can therein
split the transformation $T=\big({1\atop \tilde Z}{Z\atop 1}\big)$ ---
which generates the matrix $Q=T\Lambda T^{-1}$ out of the trivial
configuration $\Lambda$ --- into two factors,
\begin{align}
  \label{eq:multpert}
  T=T_0\,T_{\rm d}\,,\qquad
  T_0=\left({1\atop \tilde B}{B\atop 1}\right)\mathbb{1}_{\rm QD}\,,\qquad
  T_{\rm d}=\left({1\atop \tilde Z_{\rm d}}{Z_{\rm d}\atop  1}\right)\,.
\end{align}
The first factor, $T_0$, makes up the zero dimensional sigma
model while $T_{\rm d}$ comprises all non-universal configurations.

The two options can be made equivalent. To that end we observe that
the product $T_0T_{\rm d}=\left({1+B\tilde Z_{\rm d}\atop \tilde
    B+\tilde Z_{\rm d}}{B+Z_{\rm d}\atop 1+\tilde BZ_{\rm d}}\right)$
has diagonal entries different from unity and thus seems to miss the
required structure. We need not worry, however, since we may split off
a block diagonal right factor, $T_0T_{\rm d}=T'V$ with
$V=\left({1+B\tilde Z_{\rm d}\atop }{\atop 1+\tilde BZ_{\rm
      d}}\right)$, such that $T'=\left({1\atop {(\tilde B+\tilde
      Z_{\rm d})}{1\over 1+ B \tilde Z_{\rm d}}} {(B+Z_{\rm d}){1\over
      1+\tilde B Z_{\rm d}}\atop 1}\right)$ does have unit diagonal
entries. The block diagonal $V$, on the other hand, cancels from the
action since it commutes with $\Lambda$. In other words, the product
of two good transformations $T$ is good again (see
Appendix \ref{app:Qinvariance}). We conclude that the two options
of capturing deviations from the zero dimensional sigma model are
equivalent if we stipulate
\begin{align}
  \label{eq:dvsdelta}
  B+\delta Z=(B+Z_{\rm d}){1\over  1+\tilde B Z_{\rm d}}\quad\qquad
  \mathrm{and} \qquad\quad
  \tilde B+\delta \tilde Z=(\tilde B+\tilde Z_{\rm d}){1\over  1+ B \tilde Z_{\rm d}}\,.
\end{align}

For reasons to be revealed presently, we now choose $Z_{\rm d},\tilde
Z_{\rm d}$ (rather than $\delta Z, \delta \tilde Z$) as basic
representatives of 'fluctuating configurations' of $Z,\tilde
Z$. Independence of the mean fields $\mathbb{1}_{\mathrm{QD}}\,B,
\mathbb{1}_{\mathrm{QD}} \,\tilde B$ and the fluctuations $Z_{\rm
  d},\,\tilde Z_{\rm d}$ is secured by requiring the latter to be
traceless in QD,
\begin{align}
  \label{eq:traceless}
  {\rm tr_{QD}}Z_{\rm d}={\rm tr_{QD}}\tilde Z_{\rm d}=0\,.
\end{align}
We shall expand the action in powers of the fluctuating fields $Z_{\rm
  d},\tilde Z_{\rm d}$. Starting with in the form (\ref{eq:11}) for
$\mathcal S$ we set $Z=B+\delta Z$ with $\delta Z$ expressed in terms
of $Z_{\rm d}$ as stipulated above in (\ref{eq:dvsdelta}). Of course,
the latter equivalence relation must be expanded as well, $\delta Z=
(1-B\tilde B)\big(Z_{\rm d}-Z_{\rm d}\tilde B Z_{\rm d}\big)+\ldots$
and similarly for $\delta \tilde Z, \tilde Z_{\rm d}$\,. The reason we
use the form (\ref{eq:11}) rather than (\ref{eq:susySwithQ},\ref{eq:Q})
as a starting point is a considerable saving of labor and space.

Why prefer $Z_{\rm d},\tilde Z_{\rm d}$ and thus insisting in the
compatibility of the perturbation expansion to be expounded with the
representation (\ref{eq:susySwithQ},\ref{eq:Q}) of the action?  The
principal reason for our preference is that we so preserve the
manifold (a symmetric space, see Appendix \ref{app:symspace}) on which
the supermatix $Q$ lives. As a practical benefit of that choice we
shall presently find our perturbation expansion, formally one in
powers of the fluctuations, to actually go in powers of small
parameters. Those latter will turn out the three phases ${e\over
  N},{\epsilon_\pm\over N}$ representing the three combinations of
$a,b,c,d$ which the generating function has as its independent
variables, cf.~(\ref{eq:3combs}).

We shall be led to speaking of the fluctuating configurations $Z_{\rm
  d},\tilde Z_{\rm d}$ as 'decaying modes' and would like to motivate
that manner of speaking now. As already mentioned, it will
prove helpful to choose a QD representation in which the fluctuations
acquire a semiclassical meaning in the limit $\hbar\to0$. Both the Husimi
and Wigner function are suitable. Then the single-step Floquet map
becomes, up to corrections vanishing as $\hbar\to0$, the classical
Frobenius-Perron map under which the 'traceless' fluctuation $\tilde
Z_{\rm d}$ indeed decays, with rates known as resonances \cite{Gaspa98}.

\subsection{Formal expansion and identification
    of small parameters}
\label{subsec:expansion}

We start with decomposing the action in three additive pieces,
$\mathcal S=N \mathcal S_0+\mathcal S_{\rm d}+\mathcal S_{\rm c}$. The
first, $N \mathcal S_0=\mathcal S|_{Z_{\rm d}=\tilde Z_{\rm d}=0}$,
makes up the zero dimensional sigma model; the second, $\mathcal
S_{\rm d}=\mathcal S|_{B=\tilde B=0}$, pertains to the free decaying
modes, while the remainder $\mathcal S_{\rm c}$ refers to the
coupling,
\begin{align}
  \nonumber
   \mathcal S_{\rm c}=
&\;-\mathrm{str}\left({1\over 1-\tilde B\hat B}\,(1-\tilde BB)U\tilde Z_{\rm d}U^\dagger
\hat e_-(1-\tilde B B)Z_{\rm d}\hat e_+\right)
+\,\mathrm{str}\,U\tilde Z_{\rm d}U^\dagger\hat e_-Z_{\rm d}\hat e_+
\\ \nonumber &
-\mathrm{str}\left({1\over 1-\tilde B\hat B}\,(1-\tilde BB)\,U\tilde Z_{\rm d}U^\dagger
\hat B\,{1\over 1-\tilde B\hat B}\,\tilde B\hat e_-\,(1-B\tilde B)
Z_{\rm d}\hat e_+\right) +\ldots\\ \label{eq:couplingaction}
=&\;\mathrm{str}\,U\tilde Z_{\rm d}U^\dagger\hat e_-Z_{\rm d}\hat e_+
-\mathrm{str}\left({1\over 1-\tilde B\hat B}\,(1-\tilde BB)\,U\tilde Z_{\rm d}U^\dagger
\,{1\over 1-\hat B\tilde B}\,\hat e_-\,(1-B\tilde B)Z_{\rm d}\hat e_+\right)
+\ldots\,
\end{align}
with the shorthand $\hat B=\hat e_-B\hat e_+$. We have not bothered to
write out the terms linear in $Z_{\rm d}$ and $\tilde Z_{\rm d}$ (they
vanish due to the tracelessness in QD) as well as the ones $\propto
Z_{\rm d}^2,\,\tilde Z_{\rm d}^2$ (they vanish in the mean coupling
$\langle \mathcal S_{\rm c}\rangle$). In fact, on display are only the
bilinear terms $\propto Z_{\rm d}\tilde Z_{\rm d}$; higher-order terms
would have to be accounted for in higher orders of the perturbation
expansion which it will not be necessary to go into. Now, \textit{our
  coupling action vanishes as $\hat e_\pm\to 1$; that fact allows to
  argue that the small parameters underlying the expansion are the
  phases ${e\over N},\,{\epsilon_\pm\over N}$}.

We are indeed free to take as small the phases $\epsilon_\pm$ (or,
equivalently, the differences $a-c$ and $b-d$), since these source
variables just serve to get the correlator from the generating
function according to (\ref{eq:CfromZ}). We shall therefore save labor
by carrying our perturbation expansion only as far as necessary to
capture the term $\propto(a-c)(b-d)\propto \epsilon_+\epsilon_-$. On
the other hand, assuming smallness of ${e\over N}$ makes for a
non-trivial restriction. Our perturbative procedure cannot cover the
phase scale $e\sim N$ on which the correlator manifests, by its
definition, periodicity with period $N\pi$.  However, the range of
correlation decay wherein the correlator is of non-negligible
magnitude, say $C(e)\gg{1\over N^2}$, is fully contained. We are not
even limited to ${e\over N}$ of order ${1\over N}$; allowable is
$e\sim N^{\alpha}$ with $\alpha <1$.  If one wants to restrict $e$ to
the range where the correlator $C_{\mathrm{CUE}}$ is larger than any
system specific noise the exponent $\alpha$ must be smaller yet, see
\cite{Braun14}. For instance, if smoothing is done by Im\,$e$ small
compared to unity and independent of $N$ one has $\alpha<{1\over
  4}$. With that understanding we can regard ${e\over N}$ as the small
parameter making our perturbation expansion meaningful.

We should not silently pass over the question whether or not an
expansion in powers of ${e\over N}$ can capture the behavior of $R(e)$
on the Ehrenfest scale $e_E\sim {N\over \ln N}$.  Inasmuch as on that
scale ${e\over N}$ is still small, ${e_E\over N}\sim{1\over \ln N}$,
one might hope for the expansion to remain marginally
sensible. However, we do not indulge since any such correction is
drowned by system specific noise.

The free-decay piece of the action becomes
\begin{align}
  \label{eq:28}
  \mathcal S_{\rm d}=\mathrm{str}\,(\tilde Z_{\rm d}Z_{\rm d}
                          -U\tilde Z_{\rm d}U^\dagger \hat e_-Z_{\rm d}\hat e_+)\,;
\end{align}
again, we include only the leading contribution $\propto Z_{\rm
  d}\tilde Z_{\rm d}$. Terms proportional to both $Z_{\rm d}^n$ and
$\tilde Z_{\rm d}^n$ with $n>1$ would accompany similar terms in
$\mathcal S_{\rm c}$ in higher orders.  We now represent the decaying
modes by Wigner functions $Z_{\rm d}(x),\,\tilde Z_{\rm d}(x)$, with
$x$ designating the pertinent phase-space variables. Then the
tracelessness (\ref{eq:traceless}) is  expressed by phase-space
integrals, $\int dx Z_{\rm d}(x)=\int dx \tilde Z_{\rm d}(x)=0$, and
the QD trace of a product of $Z_{\rm d}$ and $\tilde Z_{\rm d}$
becomes the integral ${\rm tr_{QD}}A\tilde Z_{\rm d}BZ_{\rm d}=\int
\!dx A\tilde Z_{\rm d}(x)BZ_{\rm d}(x)$ with $A,B$ arbitrary BF
matrices. The integration range for all $x$ integrals is the phase
space volume $\int\! dx =\Omega$. The stroboscopic time evolution can
be accounted for by a propagator $\mathcal F$,
\begin{align}
  \label{eq:22}
  U \tilde Z_{\rm d}U^\dagger\to \mathcal F\tilde Z_{\rm d}(x)\,.
\end{align}

The stage is thus set for implementing the semiclassical limit. To
within corrections of higher than first order in $\hbar$ the Wigner
function propagator becomes the classical Frobenius-Perron operator
\cite{Schle11}. Different types of classical decay are possible
\cite{Gaspa98}. It may here suffice to look at the case of purely
exponential decay, with complex frequencies $\lambda_\mu$ called
Pollicott-Ruelle resonances. These come with normalizable biorthogonal
right and left eigenfunctions of $\mathcal F$, to be denoted as
$\tilde c_\mu(x)$ (right) and $c_\mu(x)$ (left). The eigenvalues of
$\mathcal F$ are $\e^{-\lambda_\mu}$. We expand as $\tilde Z_{\rm d}
(x)=\sum_\mu \tilde z_\mu \tilde c_\mu(x)$, the latter sum excluding
the ergodic stationary eigenfunction of $\mathcal F$.  The free-decay
and the coupling parts of the action then become sums over resonances,
\begin{align}
  \label{eq:Scmu}
  \mathcal S_{\rm c}&=\sum_\mu \mathcal S_{\rm c\mu}=
                             \sum_\mu\e^{-\lambda_\mu}\mathrm{str}
\Big(\,\tilde  z_\mu \hat e_-z_\mu\hat e_+
-{1\over 1-\tilde B\hat B}\,(1-\tilde BB)\,\tilde z_\mu
{1\over 1-\hat B\tilde B}\,\hat e_-\,(1-B\tilde B)z_\mu\hat e_+
\Big)\,,
\\ \label{eq:Sdmu}
   \mathcal S_{\rm d}&=\sum_\mu \mathcal S_{\rm d\mu}
               =\mathrm{str} \Big(\tilde z_\mu z_\mu
                  -\e^{-\lambda_\mu}\tilde z_\mu \hat e_-z_\mu\hat e_+\Big)\,;
\end{align}
here and below the supertrace refers to BF only.

Turning to the generating function $\cal Z$ given by the
integral (\ref{eq:10}) we see the integration variables fall into
independent subsets pertaining to $B,\,\tilde B,Z_{\rm d},\,\tilde
Z_{\rm d}$. Expanding the integrand in powers of the coupling and
dropping terms of higher than first order in $\mathcal S_{\rm c}$ we
get
\begin{align}
  \label{eq:finalZ}
  \mathcal Z=\int d(B,\tilde B)\, \e^{-N\mathcal S_0}
                    \int d(Z_{\rm d},\tilde Z_{\rm d})\, \e^{-\mathcal S_{\rm d}}
                    \big(1-\mathcal S_{\rm c}+\ldots\big)
                  \equiv\mathcal Z_0\mathcal Z_{\mathrm{d}}
                    \big(1-\langle \mathcal S_{\rm c}\rangle+\ldots\big)
\end{align}
where $\mathcal Z_0$ is the generating function of the zero
dimensional sigma model evaluated in the previous section, $\mathcal
Z_{\rm{d}}$ the factor exclusively contributed by the decaying modes,
\begin{align}
  \label{eq:Zd}
  {\cal Z}_{\rm d}= \prod_\mu \bigg(1+{\e^{-\lambda_\mu}(a-c)(b-d)\over
         (1-ab\e^{-\lambda_\mu}) (1-cd\e^{-\lambda_\mu})}\bigg)\,,
\end{align}
and $\langle \mathcal S_{\rm c}\rangle$ the relative correction due to
the coupling.  Like the full generating function $\mathcal Z$ and the
mean-field one $\mathcal Z_0$, the free-decay part $\mathcal Z_{\rm d}$
is now seen to deviate from unity by a summand proportional to
$(a-c)(b-d)$. The mean coupling must therefore obey
\begin{align}
  \label{eq:meancoupling}
  \langle\mathcal S_{\rm c}\rangle\propto (a-c)(b-d).
\end{align}

We can now harvest the fruits of the perturbative labor for the
correlator $C(e)$, using (\ref{eq:CfromZ}). The forgoing results for
the generating function entail
\begin{align}
  \label{eq:finalC}
  C(e)=C_0(e) + \textstyle{2 \e^{\I 2e/N}\over N^2}
             \partial_c\partial_d\mathcal Z_d\big|_{\epsilon_{\pm}=0}
          -\textstyle{2 \e^{\I 2e/N}\over N^2}\partial_c\partial_d
            \langle \mathcal S_{\rm c}\rangle\big|_{\epsilon_{\pm}=0}
\end{align}
where we have used $\langle \mathcal S_{\rm
  c}\rangle\big|_{\epsilon_{\pm}=0}=0$.  The first of the two
corrections to the universal form $C_0(e)$ is fixed by (\ref{eq:Zd})
as the sum over resonances
\begin{align}
  \label{eq:168}
  \textstyle{2 \e^{\I 2e/N}\over N^2}
   \partial_c\partial_d\mathcal Z_d\big|_{\epsilon_{\pm}=0}=
  {2\over N^2}\sum_\mu{\e^{\I 2e/N-\lambda_\mu}\over (1-\e^{\I 2e/N-\lambda_\mu})^2}
  \,.
\end{align}
The existence of a finite gap
\begin{equation}
\label{eq:gap}
  \Delta_g= \mathrm{min}_{\lambda_\mu \neq 0} \,|1-\e^{-\lambda_\mu}|>0
\end{equation}
is thus an indispensable requirement on the underlying classical
dynamics. To further appreciate the correction in question we ask
how many of the possibly infinitly many classical resonances can admitted as
contributors? While a precise answer to that
question is hard to give we can, for an order-of-magnitude argument,
certainly say that the classical phase space consist of $N$ Planck
cells. The classical approximation for the single-step quantum
evolution (\ref{eq:22}) can be trusted only inasmuch as it 'does not
try' to resolve sub-Planck-cell structures. Now, the 
resonances come with 'eigenfunctions' whose typical phase-space scales
decrease as $|\e^{-\lambda_\mu}|$ becomes smaller
\cite{Webe00,Webe01,Mande01}. Admittable resonances are the ones associated
with phase-space structures 'wider' than a Planck cell. It is hard to
imagine their number to be larger than of the order $N$. Were we to take
that imagination at face value we could estimate
an upper limit to the modulus of our above correction as the number
$N$ times the maximal modulus (over all $\mu$), thus arriving at the 'bound' 
${2\over N\Delta_g^2}\propto {1\over N}$. We hurry to add that we do
not face a bound in any strict sense since a 'common-sense' argument
is at its base; for the quantum graphs to be treated in
\ref{sec:graphs} stronger statements will be possible. ---
That estimate could in some cases be unnessecarily restrictive. For
instance, the moduli $|\e^{-\lambda_\mu}|$ (ordered such as to decrease
for growing $\mu$)  may decrease so rapidly that the sum over $\mu$ in
the correction (\ref{eq:168}) remains finite for $N\to\infty$, and for
such systems the correction will be of the order ${1\over N^2}$.

The same large-$N$ behavior is met with in the second correction term
in (\ref{eq:finalC}).  Since no new ideas are needed to check this we
defer the somewhat lengthy investigation to App.~\ref{app:lastcorr}.

We do not see the ratio ${e\over N}$ in the leading-order correction
because of the explicit factor ${1\over N^2}$ in the relation
(\ref{eq:CfromZ}) between generation function and correlator. We want
to recall that our perturbative treatment is confined to the windows
of correlation decay and revival, where ${e \,\mathrm{mod} N\pi\over
  N}\to 0$ for $N\to\infty$; there, the CUE and GUE correlators
agree. Given the ${1\over N^2}$ corrections we conclude, as
in the numerical investigation of Ref.~\cite{Braun14}, that CUE
spectral fluctuations can, for an individual spectrum, be
distinguished from GUE ones only by the periodicity of the CUE
correlator.

We would like to underscore that we have introduced the 'ballistic
sigma model' when replacing the quantum propagator $\mathcal F$ with
the classical Perron-Frobenius operator.  To bar all possibility of
misunderstanding we note that our classical approximation for the
\textit{single-step propagation} does not amount to any prejudice for
the difference between classical and quantum dynamics at large
times. We are simply not led to iterations of the map (like $U^n\tilde
Z (U^\dagger)^n,\, n=2,3,\ldots$) which would no later than on the
Ehrenfest scale $n_E=\lambda^{-1}\ln N$ defy classical
approximation. As is well known, the ensuing effective equilibration
looks rather different in classical and quantum perspective, the never
ending classical stretching, squeezing, and folding being washed out
by quantum fluctuations \cite{Altla12}. Moreover, inasmuch as the
decaying fields $Z_{\mathrm d}$ and $\tilde Z_{\rm d}$ enter the
action pieces (\ref{eq:couplingaction},\ref{eq:28}) only bilinearly we
are allowed to write the QD trace of the product $Z_{\mathrm d}\tilde
Z_{\rm d}$ as a phase-space integral over the product of the Wigner
functions. Higher orders of the perturbation expansion would, of
course, bring in Moyal products for more than two Wigner functions
\cite{Muella07}.

\section{Sigma model of autonomous flows via stroboscopic
  description}
\label{sec:flows}

The color-flavor transformation employed in Sect.~\ref{sec:susymaps}
for periodically driven systems can also be made availabe for
autonomous flows, simply by resorting to a stroboscopic description
with a suitably chosen strobe period $\tau$ and the pertinent Floquet
operator $U=\e^{-\I H\tau/\hbar}$. Such a stroboscopic description of
autonomous flows has already proven fruitful for the semiclassical
treatment of spectral fluctuations \cite{Braun12b}.

We consider a (sub)spectrum of $N$ consecutive energy levels $E_1\leq
E_2\leq\ldots \leq E_N$ and the pertinent unimodular eigenvalues of
$U$, $\e^{-\I E_m\tau/\hbar}$,\,$m=1,2,\ldots N$. The strobe period
$\tau$ is then chosen such that the eigenphases alias quasienergies
$\phi_m=E_m\tau/\hbar$ fill the $2\pi$ interval just once. The first
and last energy levels thus become nearest-neighbor quasienergies
whose angular distance can be fixed as the mean spacing $2\pi/N$. In
the limit of large $N$ the artificial nearest-neighbor relationship of
$\phi_1$ and $\phi_N$ looses any significance. Inasmuch as the strobe
period can be considered a classical quantity independent of Planck's
constant the extent of the spectrum becomes of order $\hbar$,
i.~e.~$E_N-E_1=\mathcal O(\hbar)$.

With the Floquet operator thus defined fed into the generating
function (\ref{eq:1}) all considerations of the preceeding sections
run through unchanged. Of course, the periodicity in the phases is now
an artefact to be shed by changing variables as (\ref{eq:3combs}) and
going to the limit $N\to\infty$ at constant $e$.

The stroboscopic description of autonomous flows and the ensuing
construction of the 'ballistic' sigma model through the color-flavor
transformation offers advantages. Neither Hubbard-Stratonovich
transformation, nor saddle-point approximation, nor regularization by
noise are needed, in contrast to previous versions of the ballistic
sigma model \cite{Musyk95,Andre96,Andre96a}. Such advantages
notwithstanding, it is appropriate to acknowledge that the transition
to the ballistic limit is very similar now and in the previous
treatments: where we replace the single-step quantum propagation
$U\tilde Z U^\dagger$ by the classical (Fobenius-Perron) propagation
of the Wigner function $\tilde Z(x)$, the previous procedure for
autonomous flows was the same for an infinitesimal time interval,
i.~e., to replace ${\I\over\hbar}[H,\tilde Z]$ by the action of the
classical Liouville operator on the Wigner function. Of course, in
every other respect, the quantum character of the 'fields' $Z$ and
$\tilde Z$ must be retained; as already mentioned, products of more
than two such fields are to be understood as Moyal products
\cite{Muella07}. It just so happened that we did not need to go to
products of more than two fields in Sect.~\ref{sec:validityof0}.

\section{Kicked rotor, quantum localization}
\label{sec:rotor}
\subsection{Introductory remarks}

Quantum localization for the kicked rotor ('Chirikov's standard
model') is nearly as old a topic as the whole field of quantum chaos
\cite{Chiri86,Casat86,Shepe86,Chiri08}. Even the field theoretic
treatment in terms of the ballistic sigma model was introduced more
than a decade ago \cite{Altla96}.  Strong evidence for the power of
the sigma model has come from the recent extension recovering
everything previously known about the rotor and even including higher
dimensional cases \cite{Tian10,Tian11}.

Our goal in this review is a comparatively modest one. Focussing
on the simplest situation (a single classical degree of freedom,
practically full chaos in the classical phase space, unitary symmetry
class, absence of quantum resonances) we propose to develop the
pertinent sigma model along a pedestrian path which should not be hard
to follow for readers previously uninitiated.

Arbitrarily long lived excitations (slow diffusion of the angular
momentum) then arise classically, forbidding a finite gap in the
Frobenius-Perron spectrum.  The quantum signature of the slow
diffusion is localization (of the Floquet eigenfunctions in the
angular momentum representation). The quasi-energy spectrum has
Poissonian rather than CUE statistics, simply because Floquet
eigenfunctions without overlap in angular momentum space provide no
reason for their quasi-energies to repel.

One can start with a time dependent Hamiltonian,
$\hat{H}(t)= {(\hbar \hat n)^2\over 2I} + k \cos(\hat{\theta}+a)
\sum_m \delta(t-mT)$\,, where $\hat{\theta}$ and $\hbar
\hat{n}=-\I\hbar\partial_\theta$ are operators for the angular
coordinate and angular momentum, obeying the canonical commutation
relation $[\hat{\theta},\hat{n}]=\I$; the momentum of inertia is
denoted by $I$ and $T$ is the kicking period.  A finite value of the
phase $a$ breaks invariance under the time reversal transformation
$t,\theta,l\mapsto-t,-\theta,l$.  We shall in the following employ the
Floquet operator describing the time evolution over one period
$T$ (from right before a kick to right before the next),
\begin{align}
\label{eq:u}
\hat{U}
=\Big(\e^{-\I\int_{0^-}^{T^-}dt H(t)/\hbar}\Big)_{\!+}=\;
\e^{-{\I\tilde{h}\hat{n}^2\over 2}}\,
\e^{-{\I K\over \tilde{h}}\cos(\hat{\theta}+a)} \,,
\end{align}
with the dimensionless representative of Planck's constant $\tilde
h={\hbar T\over I}$ and the dimensionless kicking strength $K={kT\over
  I}$. Of special interest for us is $K\gg 1$ since that limit
produces strong chaos. To exclude the resonances mentioned before we
can require $\tilde h/4\pi$ to be an irrational number. Equally
acceptable and in a sense preferable is 'near irrationality', i.e.,
$\tilde h=4\pi {M\over N}$ with $M,N$ both large primes. That
latter case comes with the conservation laws $[T_N,U]=0$ where $T_N$
is a translation in angular momentum, $T_N|n\rangle=|n+N\rangle$ with
$\hat n|n\rangle=n|n\rangle$ and integer $n$. We can then work in a
Hilbert space of dimension $N$ (the first Brillouin zone, in solid
state language, see \cite{Tian10}). Grosso modo, the physics of
localization will be unchanged if we assume $N$ extremely large
compared to the localization length.

For later use we note the matrix elements of the Floquet operator in
the angular momentum representation,
\begin{align}
\label{eq:Unm}
U_{nm}=\langle n|\hat{U}|m\rangle
         =\e^{-{\I\tilde{\hbar} n^2\over2}}\int_0^{2\pi} {d\theta\over 2\pi}\;
           \e^{-\I{K\over \tilde{\hbar}}\cos(\theta+a)+\I (n-m)\theta}\,.
\end{align}

\subsection{Inverse participation ratio}
\label{IPR}

A convenient indicator of localization is the so called inverse
participation ratio (IPR), a quantity intimately related to the
localization length (the typical span over which an eigenfunction of
$U$ is extended in angular momentum space). The convenience of the IPR
lies in the fact that it is a two-point quantity like the complex
correlator $C(e)$ such that it is accessible through the previously
used generating function, at least after slight technical
modification, and in just that way the sigma model comes into play.

For an inidividual eigenvector $|\mu\rangle$ of $U$ the inverse participation ratio
is defined as $P_\mu=\sum_{n=1}^N|\langle n|\mu\rangle|^4$ where
$\{|n\rangle,\,n=0,1,2,\ldots N\}$ are the angular momentum eigenstates
which form the basis wherein we look for localization. Now, an
extended eigenvector with $|\langle n|\mu\rangle|\sim {1\over\sqrt N}$
would have $P_\mu\sim{1\over N}$ and thus $P_\mu\to 0$ as
$N\to\infty$. A finite IPR results, however, for an exponentially
localized vector of width $\l_\mu$ independent of $N$, namely
$P_\mu\propto{1\over l_\mu}$. In order to characterize the whole
Floquet matrix rather than a single eigenvector, we employ the
spectrally averaged IPR
\begin{align}
  \label{eq:15}
  P={1\over N}\sum_{\mu=1}^N \sum_{n=1}^N|\langle n|\mu\rangle|^4\,.
\end{align}

It is easy to check that the spectrally averaged IPR is indeed related
to a two-point quantity. To that end we employ the matrix elements
$G^{\pm}_{nn}(a,\phi)=\langle n|{1\over 1-a\e^{\pm\I\phi}U^{\pm
    1}}|n\rangle$ of the retarded/advanced Green function with
$|a|=\e^{-\epsilon}$ at small positive $\epsilon$.  We propose to
scrutinize the center-phase average of the product of these two matrix
elements multiplied with ${2\epsilon\over N}$,
\begin{align}
  \label{eq:16}
  {2\epsilon\over N}\int_0^{2\pi} {d\phi\over 2\pi}
                                G^{+}_{nn}(a,\phi) G^{-}_{nn}(a,\phi)\,,
\end{align}
in the limit as $\epsilon$ approaches 0. As long as $\epsilon>0$ we
can expand both Green functions in geometric series and afterwards do
the $\phi$-integral. Resumming the resulting new geometric series we
get
\begin{align}
  \label{eq:26}
  {2\epsilon\over N}\int_0^{2\pi} {d\phi\over 2\pi}
  G^{+}_{nn}(a,\phi) G^{-}_{nn}(a,\phi)={1\over N}\sum_{\mu,\nu=1}^N
  |\langle n|\mu\rangle|^2 |\langle n|\nu\rangle|^2
  {2\epsilon\over 1-a^2\e^{\I(\phi_\nu-\phi_\mu)}}\,.
\end{align}
We are free to fix the source variable as $a=\e^{-\epsilon}$. Then
only the diagonal terms of the double sum over Floquet eigenvectors
survive the limit $\epsilon\downarrow 0$ and we indeed arrive at the
spectrally averaged IPR after summing over $n$,
\begin{align}
  \label{eq:27}
  P=\lim_{\epsilon\downarrow 0}\sum_{n=1}^N
        {2\epsilon\over N}\int_0^{2\pi} {d\phi\over 2\pi}
  G^{+}_{nn}(\e^{-\epsilon},\phi) G^{-}_{nn}(\e^{-\epsilon},\phi)\,.
\end{align}

\subsection{Towards the sigma model}

We can capture the center-phase average of the product
$G_{nn}^{+}G_{nn}^{-}$ by a superintegral in the fashion of
(\ref{eq:2}),
\begin{eqnarray}
  \label{eq:30}
\int_0^{2\pi} {d\phi\over 2\pi}G^{+}_{nn}(a,\phi) G^{-}_{nn}(a,\phi)
= \int_0^{2\pi}{d\phi\over 2\pi}\;\int d(\psi,\psi^*)
\psi_{+,B,n}^*\psi_{+,B,n}\psi_{-,B,n}^*\psi_{-,B,n}\qquad\qquad\qquad
\qquad\qquad\qquad\qquad&\\ \nonumber
\times
\exp\sum_{k,l=1}^N
\Big\{-\psi_{+,B,k}^*(\delta_{kl}-a\e^{\I\phi}\,U_{kl})\psi_{+,B,l}
             -\psi_{-,B,k}^*(\delta_{kl}-a\e^{-\I\phi}\,U_{lk}^*)\psi_{-,B,l}&
\\ \nonumber
            -\,\psi_{+,F,k}^*\,(\delta_{kl}-a\e^{\I\phi}\,U_{kl})\psi_{+,F,l}
             -\psi_{-,F,k}^*\,(\delta_{kl}-a\e^{-\I\phi}\,U_{lk}^*)\psi_{-,F,l}&
\;\Big\}\,.
\end{eqnarray}
Note that we could simplify, relative to (\ref{eq:2}), by setting
$a=b=c=d$. Indeed, then, the Gaussian integrals give just the left
side of the foregoing equation.

As in Sect.~\ref{sec:susymaps} we can apply the color-flavor
transformation (\ref{eq:6}) to trade the center-phase
average against an integral over supermatrices $Z,\tilde Z$. Instead
of (\ref{eq:7}) we now get
\begin{align}
  \label{eq:37}
  \int_0^{2\pi} {d\phi\over 2\pi}G^{+}_{nn}(a,\phi) G^{-}_{nn}(a,\phi)
  =&\int d(\psi,\psi^*)\psi_{+,B,n}^*\psi_{+,B,n}\psi_{-,B,n}^*\psi_{-,B,n}
    \exp\big(-\psi_+^{*T}\psi_+-\psi_-^{*T}\psi_-\big)\\ \nonumber &
\;\;\;\times \int d(Z,\tilde Z) \,\mathrm{sdet}(1-Z\tilde Z)
\exp\big(\psi_+^{*T}\tilde{ Z}\psi_- +
\psi_-^{*T}U^\dagger\hat e_- Z\hat e_+U\psi_+\big)\\
=&\int d(Z,\tilde Z) \,\mathrm{sdet}(1-Z\tilde Z)
\int d(\psi,\psi^*)\psi_{+,B,n}^*\psi_{+,B,n}\psi_{-,B,n}^*\psi_{-,B,n}
\,\e^{-\psi^\dagger M\psi}\nonumber
\end{align}
with the matrix $M$ given in (\ref{eq:9}). Doing the integral over
$\psi,\psi^*$ with the help of Wick's theorem we arrive at
\begin{align}
  \label{eq:46}\hspace{-.1cm}
\int_0^{2\pi} \!{d\phi\over 2\pi}G^{+}_{nn}(a,\phi) G^{-}_{nn}(a,\phi)=&\!\!
\int\!\!d(Z,\tilde Z) \,\e^{-\mathcal S}\textstyle
{\Big(\!\!\big(M^{-1}\big)_{+Bn,+Bn} \big(M^{-1}\big)_{-Bn,-Bn}
+ \big(M^{-1}\big)_{+Bn,-Bn} \big(M^{-1}\big)_{-Bn,+Bn}
\Big)}\\ \nonumber
\equiv& \int\!\!d(Z,\tilde Z) \,p_n\,\e^{-\mathcal S}\,.
\end{align}
The action $\mathcal S$ as given in (\ref{eq:11}) reappears, now with
$a=b=c=d=\e^{-\epsilon}$. We note the matrix in the prefactor $p_n$
\begin{align}
  \label{eq:48}
  M^{-1}=\left({(1-a^2U\tilde Z U^\dagger Z)^{-1}\atop
a^2Z (1-a^2U\tilde Z U^\dagger Z)^{-1}}\;\;
{U\tilde ZU^\dagger(1-a^2Z U\tilde ZU^\dagger)^{-1}\atop
(1-a^2Z U\tilde ZU^\dagger)^{-1}}\right)\,.
\end{align}
A finite value of the spectrally averaged IPV requires the foregoing
superintegral to be $\propto {1\over \epsilon}$ for
$\epsilon\downarrow 0$. Anticipating that 'divergence' to come from
the exponential factor $\e^{-\mathcal S}$ in the integrand we shall let $a\to 1$
in the prefactor.

\subsection{Slow modes and diffusive sigma model}

We  propose to simplify the action $\mathcal S$, invoking the slow
diffusive motion of the angular momentum in the classical limit (slow
on the time scale given by the kicking period). It is these slow modes
that preclude universal fluctuations in the quantum spectrum. In order
to isolate the corresponding 'soft' quantum
fluctuations we start with the action (\ref{eq:11}) which now depends only the
single source variable $a=\e^{-\epsilon}=1-\epsilon+\ldots$. Expanding
in powers of $\epsilon$ we have
\begin{align}
  \label{eq:49}
  \mathcal S=-\mathrm{str}\ln(1-\tilde ZZ)+\mathrm{str}\ln(1-U\tilde
  ZU^\dagger Z)+\epsilon \,\mathrm{str}{U\tilde ZU^\dagger Z\over 1-U\tilde
  ZU^\dagger Z}+\mathcal O(\epsilon^2).
\end{align}
Our task  simplifies considerably when we momentarily restrict the
attention to the lowest order in ($\tilde Z, Z$),
\begin{align}
  \label{eq:50}
  \mathcal S=\mathrm{str}\big(\tilde Z Z-U\tilde ZU^\dagger Z\big)
                    +\epsilon \,\mathrm{str} \,U\tilde ZU^\dagger Z+\ldots\,;
\end{align}
in the end a symmetry argument will allow us to restore the full
dependence on $Z,\tilde Z$.

Next, we employ the angular-momentum representation to write the
supertrace over BF and QD as
$\mathrm{str}(\cdot)=\mathrm{str_{BF}}\sum_n\langle
n|(\cdot)|n\rangle$. The matrix elements $Z_{nm}$ (and similarly
$\tilde Z_{nm}$) need not be treated in full generality. Rather, in
our search for slow modes we can try to neglect off-diagonal terms,
$Z_{nm}\sim\delta_{nm}Z(n)$. The following intuitive reasoning
supports that ansatz. Off-diagonal elements of $Z$ in the
angular-momentum representation carry information about the direction
of propagation in $n$-space (visible, for instance, in the
kick-to-kick behavior of the Wigner function). A momentarily
prevailing direction will be forgotten after a few kicks, simply since
the angular momentum behaves diffusively. As the diffusion proceeds
over many kicks off-diagonal elements will then indeed be
negligible. (The situation is reminiscent of the unbiased diffusion of
an electron on a 1D lattice which affords description by a density
matrix diagonal w.r.t.~the site label, on time scales larger than the
scattering time.) Moreover, the diagonal elements $Z(n)$ can be
expected to vary but slowly as the integer $n$ changes. This is
because the $n$-independent diagonal $Z$ and $\tilde Z$ commute with
$U$ and are thus strictly stationary; slowly varying diagonal elements
must thus make up the slow modes in search. To formalize the argument
we represent the slow modes by a truncated Fourier integral,
\begin{align}
  \label{eq:51}
  Z_{nm}=\delta_{nm}\int_0^{\phi_0}{d\phi\over 2\pi} Z(\phi)\e^{-\I n\phi}
\end{align}
with a cut-off $\phi_0$ to be specified presently, certainly with
$\phi_0\ll1$. We may of course imagine the
  cut-off worked into $Z(\phi)$ such that $Z(\phi) \to 0$ as $\phi$
  becomes larger than $\phi_0$. With that understanding we extend
  the $\phi$-integral in (\ref{eq:51}) and in what follows to the upper limit $2\pi$.

Given  the foregoing preparation we can write (the contribution of the
slow modes to) the action  as
\begin{align}
  \label{eq:52}
  \mathcal S=\sum_{nm}\mathrm{str_{BF}}Z(n)\tilde Z(m)\big(
                    \delta_{nm}-(1-\epsilon)|U_{nm}|^2\big)
                  =\int{d\phi\over2\pi}\int{d\phi'\over 2\pi}
                  \,k(\phi,\phi')\,\mathrm{str_{BF}}Z(\phi)\tilde Z(\phi')
\end{align}
with the kernel
\begin{align}
 k(\phi,\phi')=\sum_{nm}\left(\delta_{nm} -(1-\epsilon) |U_{nm}|^2
                                   \right) \e^{-\I(n\phi +m\phi' )}.
\end{align}
We here insert the matrix elements (\ref{eq:Unm}) of the Floquet
operator and use the identity $\sum_n\e^{\I n\phi}=2\pi\delta(\phi)$
to get
\begin{align}
k(\phi,\phi') = 2\pi \delta(\phi+\phi')\left(
                        1 - (1-\epsilon)\int_0^{2\pi}{d\theta\over 2\pi}\,
                        \e^{-\I {K\over\tilde{\hbar}} [\cos(\theta+a)-\cos(\theta+a-\phi)]} \right).
\end{align}
The parameter $a$ breaking the time reversal invariance is here seen
to disappear, simply by changing the integration variable $\theta$;
the localization length of the rotor, to be determined in what
follows, is thus independent of $a$. The restriction $0\leq \phi <
\phi_0\ll 1$ allows to expand
$\cos\theta-\cos(\theta-\phi)=-\phi\sin\theta+\ldots$. Further
restricting the cut-off as
\begin{align}\label{eq:smoothness}
\phi_0\ll {\rm max}(1,\tilde{\hbar}/K)
\end{align}
we can expand the exponential $\e^{{\I K\over\hbar}\phi\sin\theta}$ in
powers of the small quantity $K\phi\over\tilde h$,
\begin{align}
 k(\phi,\phi')
&\simeq 2\pi \delta( \phi+\phi')
 \Big(\epsilon+\left(\textstyle{K\over 2\tilde{\hbar}}\right)^2\phi^2\Big)
  \left( 1 + {\cal O}(\epsilon,\phi^2,(K\phi /\tilde{\hbar})^2) \right).
\end{align}
Finally returning to the angular-momentum representation we
approximate $\sum_n\simeq \int dn$ (recall slow fields are smooth).
The back transformation $Z(\phi)\simeq \int dn Z(n)\e^{\I n\phi}$
gives
\begin{align}
  \label{eq:54}
  \mathcal S=\int\! dn\,\mathrm{str_{BF}}\!\left(\epsilon Z(n)\tilde Z(n)
    +\Big(\textstyle{K\over 2\tilde h}\Big)^2
       \partial_nZ(n)\partial_n\tilde Z(n)\right).
\end{align}
Here, we had to preserve consistency and recall that the restriction
of $Z(\phi)$ to small $\phi$ allows to do the $\phi$-integral over the
interval $[0,2\pi]$ such that the orthogonality
$\int_0^{2\pi}{d\phi\over 2\pi}\,\e^{\I(n-m)\phi}=\delta_{nm}$
arises. The slow-mode action can be said to be small, in the following
sense: the first term is proportional to the infinitesimal $\epsilon$
while the second cannot get large due to the smoothness condition
(\ref{eq:smoothness}).

We may now invoke the promised symmetry argument
to shed the restriction to the quadratic action (\ref{eq:50}).
Employing the matrix $Q=T\Lambda T^{-1}$ with
$T=1-(\begin{smallmatrix}&Z\\ \tilde{Z} &\end{smallmatrix})$
as defined in (\ref{eq:Q}) we note that transformations $T$
infinitesimally close to unity entail $\mathrm{Str}(\partial_nQ)^2\sim
-8\,\mathrm{str}(\partial_n Z) (\partial_n \tilde Z)$ and
$\mathrm{Str}\,Q\Lambda\sim 4\,\mathrm{str}Z\tilde Z$ and thus
\begin{align}
\label{S(Q)con}
\mathcal S=&\,{1\over 4}\int\!dn\,\mathrm{Str}
\left(\epsilon Q(n)\Lambda-\textstyle{1\over 2}
\Big({K\over 2\tilde h}\Big)^2 \partial_nQ(n)\partial_nQ(n)
\right)
={1\over 4}\sum_n\,\mathrm{Str}
\left(\epsilon Q(n)\Lambda-\textstyle{1\over 2}
\Big({K\over 2\tilde h}\Big)^2 \Big(Q(n+1)-Q(n)\Big)^2
\right),
\end{align}
where the supertrace now is over AR and BF. At this point we may drop
the restriction of the transformation, arguing as follows. The
original matrix $Q=T\Lambda T^{-1}$ allows for general transformations
$T=1-\left(\begin{smallmatrix}&Z\\ \tilde{Z}
    &\end{smallmatrix}\right)$ with $\tilde Z_{BB}=
Z_{BB}^\dagger,\;\tilde Z_{FF}=-Z_{FF}^\dagger$ and
$|Z_{BB}Z_{BB}^\dagger|<1$ as the only conditions. In the present
context where each 'site' $n$ has its separate $Q(n)$ and we can
demand $Q(n)=T(n)\Lambda T^{-1}(n)$ with separate transformations
$T(n)=1-\left(\begin{smallmatrix}&Z(n)\\ \tilde{Z}(n)
    &\end{smallmatrix}\right)$. These single-site transformations
$T(n)$ need not be close to the identity. The respective matrices $Z(n)$ and
$\tilde Z(n)$, all $2\times 2$ in BF, are restricted as before except
that $Z_{BB}$ and $Z_{FF}$ are now complex numbers.  Letting the
latter numbers range freely otherwise we are no longer confined to
small $Z,\tilde Z$. We have arrived at what is called the (action of
the) diffusive 1D sigma model.

It remains to scrutinize the prefactor $p_n$ of $\e^{-\mathcal S}$ in
(\ref{eq:46}). As already mentioned above we may put
$a=\e^{-\epsilon}\to 1$ there since the overall factor $\epsilon$ in
the IPR cannot be compensated by the prefactor. Moreover, the slow
modes are negligible since in contrast to their role in the action
they are but small corrections of relative order
$\big(K\phi_0/\tilde h\big)^2$ in $p_n$. So we can simply put $U\to
1$. The thus simplified matrix $M^{-1}$ differs from ${1\over
  2}(Q\Lambda-1)$ only by swapped diagonal AR blocks,
$(\cdot)_{++}\leftrightarrow (\cdot)_{--}$, and likewise for the
off-diagonal blocks, $(\cdot)_{+-}\leftrightarrow (\cdot)_{-+}$. That
double swap does not change the prefactor in (\ref{eq:46}) whose
structure is $(\cdot)_{++}(\cdot)_{--}+(\cdot)_{+-}(\cdot)_{-+}$
whereupon we can write
\begin{align}
  \label{eq:53}
  4p_n=-\big(Q(n)_{+B,+B}-1\big)\big(Q(n)_{-B,-B}+1\big)
          -Q(n)_{+B,-B}Q(n)_{-B,+B}\,.
\end{align}

With both the action $\mathcal S$ and the prefactor $p_n$ expressed in
terms of the single-site matrix $Q(n)$ we have constructed the
diffusive 1D sigma model for the IPR,
\begin{align}
  \label{eq:55}
  P=\lim_{\epsilon\to 0}\sum_{n=1}^N{2\epsilon\over N}\int\!d(Z,\tilde
  Z)p_n\e^{-\mathcal S}.
\end{align}

We shall not bother to retrace the further evaluation of the IPR which
is well documented in the literature
\cite{Fyodo94,Haake10}.  It is worth stating, though,
that in the limit $N\to\infty$, that is for irrational values of
$\tilde h$, the foregoing expression can hardly be imagined to
produce anything else than, up to numeric factors, $P\propto{1\over
  l}\propto({\tilde h\over K})^2$. That expectation is suggested by
the appearance of the 
$(K/\tilde h)^2$ as the single parameter
characterizing the rotor, accompanying the term
$[\partial_nQ(n)][\partial_nQ(n)]$ in the slow-mode action.

A critical remark on our somewhat cavalier construction of the
slow-mode action is in order. It is to be complemented by checking
stability against configurations outside the slow-mode 'sector'. We
shall forgo that analysis which has been performed in
Ref.~\cite{Tian10}. The 'massive' modes we have neglected here turn
out to renormalize the diffusion constant to a weak further dependence
on $\tilde h$ and $K$ (see also \cite{Chiri86}).

\section{Quantum Graphs}
\label{sec:graphs}

\subsection{Preliminary remarks}

Ever since Kottos and Smilansky derived an exact trace formula akin to
the Gutzwiller one 15 years ago \cite{Kotto99} quantum graphs have
been a paradigmatic model of quantum chaos \cite{Gnutz06}. Graphs were
the first systems for which a microscopic derivation of the universal
two-point function could be achieved without any average over system
parameters \cite{Gnutz04,Gnutz05}.  This derivation was based on the
supersymmetry method using the color-flavor transformation. The
approach was later generalized to universal wave function statistics
\cite{Gnutz08,Gnutz10}, chaotic scattering \cite{Pluha13,Pluha13a},
and spectral correlators of all orders \cite{Pluha13b}. We here focus
on directed graphs where the method finds its clearest and simplest
realization.

\subsection{Directed graphs and their spectra}

A quantum graph $G$ consists of a metric graph and a wave equation in
the form of an eigenequation for a self-adjoint operator.  We shall
allow for $V$ vertices and $N$ edges (often called bonds) such
that each edge $e$ is attached to one vertex at either side and is
assigned a length $L_e$. We only consider connected graphs which
cannot be divided into two or more unconnected subgraphs.  For each
edge a coordinate $0\le x_e\le L_e$ and a wave function
$\phi_e(x_e)$ with $e=1,2,\ldots N$ are introduced.
We adopt the first-order wave equation
\begin{equation}
  -\I\phi'_e(x_e) = k \phi_e(x_e)
  \label{eq:momentum}
\end{equation}
with local solution $\phi_e(x_e)= a_e \e^{\I k x_e}$ where $a_e$ are
(at this point) undetermined constants.  Unlike the free Schr\"odinger
equation the first-order equation (\ref{eq:momentum}) is not invariant
under switching coordinates as $x_e \mapsto \tilde{x}_e=L_e-
x_e$. Fixing a direction on each edge we can speak of incoming and
outgoing edges for each vertex.  A self-adjoint momentum operator on a
directed metric graph can be defined \cite{Carlso99} if each vertex has
as many incoming as outgoing edges and if each vertex is assigned a
unitary scattering matrix that expresses the amplitudes of the wave
functions of the outgoing edge as a linear superposition of the
corresponding incoming amplitudes. It is then straightforward to see
that solutions to the wave equation \eqref{eq:momentum} with the above
matching conditions only exist for a discrete set of wave numbers --
the spectrum of the momentum operator.  One may characterize that
spectrum by the condition
\begin{equation}
  \zeta(k)=0
  \label{eq:quantization}
\end{equation}
for the characteristic function
\begin{equation}
  \zeta(k)= \det(1- U(k))\,, \qquad U(k)= S T(k)
\end{equation}
where $U(k)$, $S$, and $T(k)$ are three unitary $N \times N$ matrices
 acting on the wave amplitudes on each edge. The diagonal matrix $T(k)_{ee'}=
\delta_{ee'}\e^{\I kL_e}$ contains the phase
difference at the two ends of the edge, $\phi_e(L_e)= \e^{\I k L_e}
\phi_e(0) $, while the scattering matrix $S$ harbors all matching
conditions: if there is a vertex such that $e'$ is an incoming edge
and $e$ an outgoing edge then $S_{ee'}$ is the element of the
scattering matrix at that vertex; otherwise, if the two edges are not
connected through a vertex we have $S_{ee'}=0$ -- thus $S$ encodes all
matching conditions and also the connectivity of the graph. The matrix
$U(k)$ is called the \emph{quantum map} and contains all relevant
information of the quantum graph (connectivity, scattering amplitudes
at vertices, and phases for the transport along edges).  Via
\eqref{eq:quantization} the quantum map determines the momentum
spectrum.  The condition has the clear physical interpretation that
$k$ belongs to the spectrum if there is an eigenfunction such that
$\phi_e(0)=\sum_{e'} U(k)_{ee'} \phi_{e'}(0) =\sum_{e'}S_{ee'}
e^{ikL_{e'}}\phi_{e'}(0) $ -- i.e. if there is a stationary vector
under the quantum map.

Let $\left\{\e^{\I\theta_\ell(k)}\right\}_{\ell=1}^N$ be the set of
the unimodular eigenvalues of the quantum map. The phases
$\theta_\ell(k)$ are continuous functions of $k$.  It is easy to see
that these functions are strictly increasing $\frac{d
  \theta_\ell(k)}{dk}>0$, and that the spectrum of the graph is given
by $k_{\ell,n}$ such that $\theta_\ell(k_{\ell,n})=2\pi n$ ($n \in
\mathbb{Z}$).  Note that we are now dealing with two different
spectra: the infinite momentum spectrum of the quantum graph $\left\{
  k_{\ell, n}\right\}_{n \in \mathbb{Z}, \ell=1,\dots, N}$, and the
parametric $N$-dimensional quasi-energy spectrum $\left\{
  \theta_{\ell}(k) \right\}$ of the quantum map which depends on $k$
as a parameter.

It is in order to break time reversal invariance that we  treat
directed graphs and, in addition,   do not allow for
loops (edges starting and ending at the same vertex) or
'parallel' edges (i.e. there is at most one directed edge between
any two vertices) and  choose all edge lengths rationally independent.

To highlight the analogy of the quantum map for  directed
graphs to the Floquet maps of Sect.~\ref{sec:susymaps} we shall
speak of the $N$ dimensional habitat of $U, S,T$ as QD.

\subsection{Spectral statistics}
One can show \cite{Gnutz06} that the spectral statistics of the
quantum graph under study is equivalent to the $k$-averaged spectral
statistics of the quantum map -- in certain situations even with
complete rigour \cite{Berko10}. We shall here focus on the technically
simpler case of the $k$-averaged spectrum of the quantum map.  As in
the preceeding sections we describe two-point correlations in terms of
a generating function,
\begin{equation}
  \mathcal{Z}(a,b,c,d)=
  \left\langle \frac{\det\left(1-c U(k) \right)
      \det\left(1-d U(k)^\dagger \right)}{
      \det\left(1-a U(k) \right)
      \det\left(1-bU(k)^\dagger \right)} \right\rangle_k\,;
\end{equation}
the $k$-average  now employed,
\begin{equation}
  \langle F(k) \rangle_k:=
  \lim_{K \to \infty} \frac{1}{K} \int_{-K/2}^{K/2} F(k) dk\,,
\end{equation}
contains the center-phase average previously used in the Floquet map
setting. However, the $k$-average is much stronger than just center
phase since $\mathcal Z(a,b,c,d)$ depends on the parameter $k$ via $N$
unimodulars $\{\e^{\I k L_e}\}_{e=1}^N$. In fact, the map $k \mapsto
(\e^{\I k L_1}, \dots, \e^{\I kL_N})$ describes a 'trajectory' on an
$N$-dimensional torus, $k$ playing the role of a time. If the
trajectory is ergodic on the torus with uniform measure we may replace
the $k$-average (formally a time average) by an average over the torus
(the corresponding phase-space average). The trajectory is indeed
ergodic if the edge lengths $L_e$ ('frequencies') are rationally
independent. Assuming that scenario we can replace $kL_e\mapsto
\varphi_e$ and
\begin{align}
  \label{eq:Nfoldaverage}
  \langle (\cdot) \rangle_k\to\int {d^N\varphi\over (2\pi)^N}(\cdot) \,.
\end{align}

The generating function thus involves an $N$-fold phase average,
\begin{equation}
  \mathcal{Z}(a,b,c,d)=\int \frac{d^N {\varphi}}{(2 \pi)^N}\;
  \frac{\det\left(1-c T(\boldsymbol{\varphi}) S \right)
    \det\left(1- d S^\dagger T(-\boldsymbol{\varphi})\right)}{
    \det\left(1-a T(\boldsymbol{\varphi}) S \right)
    \det\left(1-b S^\dagger T(-\boldsymbol{\varphi}) \right)} \,,
  \label{eq:genfuncgraph}
\end{equation}
where $T(\boldsymbol{\varphi})_{ee'}= \delta_{ee'} \e^{\I\varphi_e}$ is
now a diagonal matrix of random phases.
We can even consider the matrices $T(\boldsymbol{\varphi}) S$ for a
fixed matrix $S$ as a random ensemble, and that ensemble is guaranteed
to have spectral correlations identical to the ones of the original
individual quantum graph.  Such ensembles have been called
unistochastic \cite{Tanne01}.  We also note that some system parameters
(the edge lengths) have disappeared; this is a first glimpse of
universality. In order to see universality in the sense of the BGS
conjecture we now move on and look for the conditions for the
unistochastic ensemble $T(\boldsymbol{\varphi}) S$ to be equivalent to
the CUE.

\subsection{Sigma model}

The construction of the sigma model proceeds as in
Sect.~\ref{sec:susymaps} and again leads to
\begin{align}
  \mathcal{Z}(a,b,c,d)&= \int d(Z,\tilde{Z})\,
  \mathrm{sdet}\left( 1-Z \tilde{Z} \right)
  \mathrm{sdet}^{-1} \left(1-
    S  \tilde{Z} S^\dagger \hat{e}_- Z \hat{e}_+
  \right)=  \int d(Z,\tilde{Z})\,
  \e^{ -\mathcal S(Z, \tilde{Z})}
  \label{eq:gsm}\\ \label{eq:graphaction}
  \mathcal S(Z, \tilde{Z})&= -\mathrm{str} \ln \left( 1- \tilde{Z} Z\right)+
  \mathrm{str} \ln \left(1- S\tilde{Z} S^\dagger \hat{e}_- Z \hat{e}_+
  \right)\, .
\end{align}
We see the graph scattering matrix $S$ taking the role of the Floquet
matrix $U$; the matrix $T(\bf{\varphi})$ has disappeared through the
phase average (\ref{eq:Nfoldaverage}). The only other difference to
Floquet maps which in fact amounts to a considerable simplification is
hidden in the compact notation.

For quantum maps, the supermatrices $Z,\tilde Z$ were full matrices
and the integral involved $4N^2$ real commuting (and as many
anti-commuting) degrees of freedom.  For quantum graphs, the
supermatrices $Z,\tilde Z$ are block-diagonal and we are  left
with only $4N$ real commuting (and as many anti-commuting) degrees of
freedom in the remaining integral.  The origin the simplification lies
in the $N$-fold phase average (\ref{eq:Nfoldaverage}) in the
generating function (\ref{eq:genfuncgraph}).  Each edge on the graph
has a separate phase average and thus affords its own color-flavor
transformation (single flavor) with $2\times 2$ supermatrices $Z_e$
and $\tilde Z_e$ in BF. All edges together give rise to the block
diagonal $Z_{ee'}=\delta_{ee'}Z_e$ and $\tilde
Z_{ee'}=\delta_{ee'}\tilde Z_e$.

In order see how classical dynamics sneaks in and universal behavior
arises we introduce the mean fields
\begin{align}
  B_{ss'}=  \frac{1}{N} \sum_{e=1}^N Z_{e, ss'}\,,  \qquad\qquad
  \tilde{B}_{ss'}=  \frac{1}{N} \sum_{e=1}^N \tilde{Z}_{e, ss'}
\end{align}
and deviations therefrom through
\begin{align}
  Z_{b,ss'}= B_{ss'} + \delta Z_{b, ss'}\,, \qquad\qquad
  Z_{b,ss'}=\tilde{B}_{ss'} + \delta \tilde{Z}_{b, ss'} \,.
\end{align}
Clearly, the 'mean-field approximation' which neglects the
fluctuations gives the zero dimensional sigma model.

For the same reasons as in Sect.~\ref{sec:validityof0} we now choose
the 'decaying fields' $Z_{\mathrm d}$ and $\tilde Z_{\mathrm d}$
rather than the 'fluctuations' $\delta Z$ and $\delta\tilde Z$ as
basic representatives of non-universal corrections, using the
relations (\ref{eq:dvsdelta}) and requiring the tracelessness
(\ref{eq:traceless}) of the decaying fields. The further treatment of
the non-universal corrections then parallels the one given for general
Floquet maps, with the graph specific simplifications due to the
block diagonality of $Z_{\mathrm d}$ and $\tilde Z_{\mathrm d}$. In
particular, the coupling part of the action (\ref{eq:couplingaction})
reduces to
\begin{align}
  \label{eq:89}
   \mathcal S_{\rm c}=\sum_{ee'}|S_{ee'}|^2
\;\mathrm{str} \Big(\,\tilde Z_{{\mathrm d}e}\hat e_-Z_{{\mathrm d}e'}\hat e_+
-{1\over 1-\tilde B\hat B}\,(1-\tilde BB)\,\tilde Z_{{\mathrm d}e}
\,{1\over 1-\hat B\tilde B}\,\hat e_-\,(1-B\tilde B)Z_{{\rm d}e'}\hat e_+\Big)
+\ldots\,.
\end{align}
As a most remarkable consequence of the block diagonality of the
fluctuations the quantum dynamics is represented by the edge-to-edge
transition probabilities
\begin{align}
  \label{eq:14}
  \left|S_{ee'}\right|^2=\left|U(k)_{ee'}\right|^2 =\mathcal F_{e e'}
\end{align}
which are independent of the wave number $k$. The dynamics thus
acquires a classical appearance, without approximation. (Recall that
for general Floquet maps classical dynamics resulted from the
approximation (\ref{eq:22}).)

Interestingly, the classical dynamics ruling the small fluctuations on
our connected directed graph is stochastic rather than Hamiltonian.
We are actually facing a Markov process since the matrix $\mathcal F$
is bistochastic, i.e., $\sum_e \mathcal F_{ee'}=\sum_{e'} \mathcal
F_{e e'}=1$. The uniform probability distribution $P_e= 1/N$ is
invariant under the action of the map $\mathcal F$, that is
$\sum_{e'}\mathcal F_{ee'} P_{e'}= N^{-1} \sum_{e'} \mathcal
F_{ee'}=1/N= P_{e}$.  If we assume that all processes $e \to e'$
allowed by the connectivity of the graph have some finite probability,
the map $\mathcal F$ is even ergodic, and the uniform distribution is
the unique stationary mode with its eigenvalue $1$ of $\mathcal F$
non-degenerate. All other eigenvalues $\e^{-\lambda_\mu}$ have modulus
below one, $|\e^{-\lambda_\mu}| < 1$; the possibly complex rates thus
have positive real parts, Re\,$\lambda_\mu>0$.  The time scale on
which ergodicity becomes apparent in the dynamics is given by the
inverse of the classical gap $\Delta_g$ introduced in (\ref{eq:gap})
analogous to the gap of the ergodic Frobenius-Perron operator in the
Floquet map setting (which latter, however, does not depend on
$N$). As already found for Floquet maps, the gap $ \Delta_g$ is the
central quantity that governs the strength of deviations from
universality in spectral correlations of quantum graphs. We should
note that only the decaying eigenmodes  are admitted in
$Z_{\rm d}, \tilde Z_{\rm d}$, the stationary one being
barred by the tracelessness of the decaying fields.

The final result (\ref{eq:finalC}) for the complex correlator $C(e)$
of our treatment of general Floquet maps can now be taken over but
requires a slightly refined interpretation.  The gap $\Delta_g$ need
not stay fixed and may even close for a sequence of quantum graphs (in
contrast to a Floquet map) with growing $N$. Moreover, the number of
resonances is exactly $N$ (again in contrast to a Floquet map where
the number of resonances is ignorant of $N$). In particular, the first
of the two correction terms in (\ref{eq:finalC}) can be estimated as
\begin{align}
  \label{eq:90}
  \Big|{2 \e^{\I 2e/N}\over N^2}
   \partial_c\partial_d\mathcal Z_d\big|_{\epsilon_{\pm}=0}\Big|\leq
  {2\over N}{1\over \Delta_g^2}\,.
\end{align}
To obtain the upper limit we have replaced the sum over resonances by
the factor $N$. Now a gap remaining finite gives a correction of
the order ${1\over N}$, rather than the ${1\over N^2}$ correction
found for Floquet maps (see (\ref{eq:168})). In the graph case, the
correction becomes even larger when the gap $\Delta_g$ closes as
$N\to\infty$, say according to the power law $\Delta_g = N^{- \nu}$.
The correction under discussion then vanishes asymptotically only as
long as $\nu < 1/2$. Again, we should recall that our analysis is
valid only within the windows of correlation decay and revival where
$C(e)$ remains finite for $N\to\infty$.

The second term in the correction (\ref{eq:168}) can be estimated
similarly.

\section{Open ends}
\label{sec:wrapup}

We hope to have presented a convincing case for the usefulness and
flexibility of the sigma model for unitary maps. Put in a nutshell, we
have shown in an at least self-consistent manner that (i) spectral
fluctuations (as captured in the two-point correlator of the level
density) of individual dynamics become universal for $N\to\infty$,
given full chaos and a gapped set of classical resonances; and (ii)
the absence of a gap in the resonance spectrum of the kicked rotor is
accompanied by quantum localization whose effect on wavefunction
statistics is captured in a nonlinear sigma model that includes all
soft modes.

The range of applicability is far from being exhausted in the existing
literature.  To mention but a few examples of worthwhile studies, the
kicked top and its approach to the kicked rotor in a certain limit for
its control parameters appears quite doable and so might even be
rigorous treatments of the simplest chaotic quantum
graphs. 

The  lowest-order
corrections ($\sim {1\over N^2}$ or ${1\over N}$, see
Sect.~\ref{subsec:expansion}) to the CUE spectral fluctuations
provided by the perturbative treatment of decaying modes are smaller
than the $1\over\sqrt N$ noise found numerically for individual
dynamics (kicked top). As already stated in the introduction, an analytic explanation could be found in extending our
work to the four-point correlator of the level density. We conjecture
that there, too, the decaying modes make for a negligible correction
to universal behavior which latter implies a ${1\over
  \sqrt{N\mathrm{Im}e}}$ typical deviation of the single-spectrum
two-point correlator from its CUE mean \cite{Braun14}.  For quantum
graphs, that behavior is in fact implied by results reported in
Ref.~\cite{Pluha13b}.

A caveat is in order. Before starting to work with
the sigma model for some physical system (or class of systems) one
must be sure to know all of its symmetries and set up the model
accordingly. For instance, when time reversal invariance reigns in the
sense of the orthogonal or symplectic symmetry classes, the space
harboring the matrices $Z,\tilde Z, Q, \ldots$ must be enlarged
relative to the one appropriate for the unitary symmetry class treated
in this review. Systems with other symmetries require their proper
adjustments of the model as well. A well known example of symmetries
that make for non-universal spectral fluctuations even though the
classical behavior is fully chaotic, are the Hecke symmetries making
for quantum conservation laws without classical counterparts. Finding
the proper setup for the sigma model in such cases of 'arithmetic
chaos' \cite{Bogom97} is an interesting challenge. Equally interesting
would be a sigma model treatment of the exceptional behavior of the
cat map \cite{Keati91}.

We believe that the existing literature on the kicked rotor can be a
good guide in further applications. The rotor exhibits a wealth of
behavior, like renormalization of the diffusion constant by massive
modes, different types of resonances, accelerator modes, Anderson
transitions in dimensions larger than one, which one after the other
have been captured in successive refinements of the sigma model,
accounting for the respective properties of the Floquet operator.

We would like to finally remark that for universal spectral
fluctuations, the sigma model is an alternative to Gutzwiller's
periodic-orbit theory \cite{Gutzw90}. A sum over closely packed
bunches of periodic orbits there arises which is equivalent to a
perturbation expansion of the zero dimensional sigma model
\cite{Braun12b,Muell09,Heusl07,Muell05,Muell04}. While periodic-orbit
theory retains the charm of revealing practicability and power of
Gutzwiller's semiclassical quantization of chaotic dynamics, the sigma
model can now perhaps be seen as a more economic
procedure. Incidentally, the periodic-orbit sum giving universal
spectral fluctuations for the correlator $C(e)$ neglects
\textit{noisy} contributions of other orbit sets (not from closely
packed bunches) which are dispatched as 'interfering
destructively'. The pertinent noise strength could be captured as
$\propto {1\over\sqrt N}$ by studying the variance of the
single-system correlator in a periodic-orbit expansion, similarly as
in the sigma-model approach.

\vspace{.2cm}
We gratefully acknowledge financial support by the
Sonderforschungsbereich SFBTR12 'Symmetries and Universality in
Mesoscopic Systems' of the Deutsche Forschungsgemeinschaft. We are
especially indepted to Peter Braun and Martin Zirnbauer for many
helpful discussions and moral support.
\vspace{.5cm}

\appendix
\section{Color-flavor transformation}
\label{app:cft}
Zirnbauer has first proven the color-flavor transformation
(\ref{eq:cft}) using coherent-state techniques
\cite{Zirnb96,Zirnb98,Zirnb99}. The alternative to be presented here
makes full use of the invariance of the flat measure $d(Z,\tilde Z)$
under the transformation
\begin{align}
  \label{eq:6}
  Z\to Z'=(AZ+B)(CZ+D)^{-1}\,,\qquad
  \tilde Z\to \tilde  Z'=(D \tilde Z +C)(B \tilde Z+A)^{-1}
\end{align}
with supermatrices $A,B,C,D$ constrained by invertibility of
$\big({A\atop C}{B\atop D}\big)$. On the other hand, the foregoing
transformation makes for an invariance of the manifold accommodating
the supermatrix $Q$ (see Appendix \ref{app:Qinvariance}).  The
left-hand side in the color-flavor transformation (\ref{eq:cft}) is an
integral representation of a Bessel function
\begin{align}
  \label{eq:17}
  \mathrm{lhs}=\int_0^{2\pi}\frac{d\phi}{2\pi} \,\exp
\big\{\e^{\I\phi}\Psi_1^T\Psi_{2'}+\e^{-\I\phi}\Psi_2^T\Psi_{1'}\big\}
=\sum_{n=0}^\infty{y^n\over n!^2}=J_0(2\I\sqrt y)
\qquad\mathrm{with}\qquad y=(\Psi_1^T\Psi_{2'}) (\Psi_2^T\Psi_{1'})\,.
\end{align}
The above invariance of the measure is needed to reduce the right-hand side
of (\ref{eq:cft}),
\begin{align}
  \label{eq:19}
  \mathrm{rhs}=
  \int d(Z,\tilde Z)\,\mathrm{sdet}(1-Z\tilde Z)
  \exp\{\Psi_1^T\tilde Z\Psi_{1'}+\Psi_2^T Z\Psi_{2'}\}\equiv
  \Big\langle \mathrm{sdet}(1-Z\tilde Z)
  \exp\{\Psi_1^T\tilde Z\Psi_{1'}+\Psi_2^T\tilde Z\Psi_{2'}\}\Big\rangle\,,
\end{align}
to that same Bessel function.

We start with noting that the special case $B=C=0$ of (\ref{eq:6}),
$Z'=AZD^{-1},\,\tilde Z'=D\tilde ZA^{-1}$, leaves the determinant
$\mathrm{sdet}(1-Z\tilde Z)$ unchanged. Therefore, that special
transformation can be used to simplify the accompanying exponential.
We can always find invertible supermatrices $A$ and $D$ filling up the
'Bosonic basis vector' ${\bf e}=(1,0,\ldots,0)^T$ as
\begin{align}
  \label{eq:56}
  \Psi_{2'}=D{\bf e}\,,\qquad\qquad \Psi_2^T={\bf e} A^{-1}\,.
\end{align}
Now using just such $A,D$ we first write the quantity $y$ defined in
(\ref{eq:17}) as $y=({\bf e}^TA^{-1}\Psi_{2'})(\Psi_2^TD{\bf e})$ and,
second, replace $Z\to AZD^{-1}$ and
$\tilde Z\to D\tilde Z A^{-1}$ to get
\begin{align}
  \label{eq:rhs}
   \mathrm{rhs}= \Big\langle \mathrm{sdet}(1-Z\tilde Z)
  \exp\{\Psi_1^TD\tilde ZA^{-1}\Psi_{1'}+ Z_{BB,11}\}\Big\rangle\,.
\end{align}
Here, the exponent only contains one single matrix element from the
Bose-Bose block of $Z$ while all elements of $\tilde Z$ still appear.
However, we may write $\Psi_1^TD\tilde ZA^{-1}\Psi_{1'}=yZ^*_{BB,11}+(\ldots)$
and thus
\begin{align}
  \label{eq:57}
  \mathrm{rhs}= \Big\langle \mathrm{sdet}(1-Z\tilde Z)
  \exp\{ Z_{BB,11}+yZ^*_{BB,11}+\ldots\}\Big \rangle\,;
\end{align}
we have used $\tilde Z_{BB}=Z^\dagger_{BB}$ and lumped all
elements of $\tilde Z$ except for the one in the upper left corner,
$Z^*_{BB,11}$, into $(\ldots)$. All these latter elements can in fact
be set to zero in the exponent since they cannot contribute to the
integral over $(Z,\tilde Z)$, due to the remaining freedom in choosing
the matrices $A$ and $D$. In particular, we can still set $A\to
A\e^{\I\varphi},\,D\to D\e^{\I\varphi}$ with $\varphi$ a diagonal
matrix with arbitrary real Bosonic and arbitrary Fermionic diagonal
entries. All terms in $(\ldots)$ then acquire arbitrary phase factors.
No particular value of any of these phases being favored against any
other we may average over the phases and so indeed find all of $(\ldots)$
annulled. Expanding the remaining exponential and realizing that only
powers of $|Z_{BB,11}|^2$ can survive the subsequent integral we
employ binomial formulas to get
\begin{align}
  \label{eq:58}
  \mathrm{rhs}= \sum_{n}{y^n\over n!^2}\Big\langle
  \mathrm{sdet}(1-Z\tilde Z) |Z_{BB,11}|^{2n}\Big\rangle\,.
\end{align}
For the Bessel function $J_0(2\I\sqrt y)$ to arise the identity
\begin{align}
  \label{eq:62}
  \Big\langle
  \mathrm{sdet}(1-Z\tilde Z) |Z_{BB,11}|^{2n}\Big\rangle=1
\end{align}
must hold for all naturals $n$. The invariance of the measure indeed
implies this identity, as is seen by applying the shift $Z\to
(Z-B)(1-CZ)^{-1}$ and $\tilde Z\to (\tilde Z-C)(1-B\tilde Z)^{-1}$ to
the normalization of the measure, $\Big\langle \mathrm{sdet}(1-Z\tilde
Z)\Big\rangle=1$. Some algebra (see further below) shows
\begin{align}
  \label{eq:74}
   \mathrm{sdet} \big(1-(Z-B)(1-CZ)^{-1}(\tilde Z-C)(1-B\tilde Z)^{-1}\big)
   ={\mathrm{sdet}(1-Z\tilde Z)\, \mathrm{sdet}(1-BC)\over
      \mathrm{sdet}(1-B\tilde Z) \,\mathrm{sdet}(1-CZ)}\,.
\end{align}
We conclude
\begin{align}
  \label{eq:75}
  \left\langle{\mathrm{sdet}(1-Z\tilde Z)\over
  \mathrm{sdet}(1-B\tilde Z) \,\mathrm{sdet}(1-CZ)}\right\rangle
  ={1\over \mathrm{sdet}(1-BC)}
\end{align}
and use that identity for the special case where $B$ and $C$ only contain
a single non-vanishing entry,
\begin{align}
  \label{eq:76}
  B_{BB,11}=\beta\,,\qquad\qquad C_{BB,11}=\gamma\,,
\end{align}
such that we have
\begin{align}
  \label{eq:77}
  \mathrm{sdet}(1-BC)=1-\beta\gamma \,,\qquad
  \mathrm{sdet}(1-B\tilde Z) =1-\beta Z^*_{BB,11}\,,\qquad
  \mathrm{sdet}(1-CZ)=1-\gamma Z _{BB,11}\,.
\end{align}
The identity (\ref{eq:75}) thus becomes
\begin{align}
  \label{eq:78}
 \left\langle{\mathrm{sdet}(1-Z\tilde Z)
 \over (1-\beta Z^*_{BB,11})(1-\gamma Z _{BB,11})} \right\rangle
 ={1\over 1-\beta\gamma}\,.
\end{align}
Expanding in powers of $\beta$ and $\gamma$ we confirm (\ref{eq:62})
and thus the color-flavor transformation (\ref{eq:cft}).

It remains to fill in the algebra proving (\ref{eq:74}). After
multiplication of both sides with $\mathrm{sdet}(1-B\tilde Z)$ the assertion
to be proven reads
\begin{align}
  \label{eq:79}
  \mathrm{sdet}\big(1-B\tilde Z-(Z-B)(1-CZ)^{-1}(\tilde Z-C)\big)
\stackrel{?}{=}{\mathrm{sdet}(1-Z\tilde Z)\, \mathrm{sdet}(1-BC)\over
      \mathrm{sdet}(1-C Z)}\,.
\end{align}
In the left-hand-side determinant we write $B\tilde Z=B{1-ZC\over
  1-ZC}\tilde Z$ and multiply out the numerator brackets. Then the term
$B{1\over 1-CZ}\tilde Z$ cancels such that the left-hand side becomes
\begin{align}
  \label{eq:80}
  &\,\mathrm{sdet}\bigg(1+BCZ{1\over 1-CZ}\tilde Z-Z{1\over 1-CZ}\tilde Z
   +Z{1\over 1-CZ}C-B{1\over 1-CZ}C\bigg)\\ \nonumber
 =&\,\mathrm{sdet}\bigg({1+ZC\over 1-ZC}+BC{1\over 1-ZC}Z\tilde Z-{1\over 1-ZC}Z\tilde Z
   +{1\over 1-ZC}ZC-BC{1\over 1-ZC}\bigg) \\ \nonumber
 =&\,\mathrm{sdet}\bigg({1\over 1-ZC}+BC{1\over 1-ZC}Z\tilde Z-{1\over 1-ZC}Z\tilde Z -BC{1\over 1-ZC}\bigg) \\ \nonumber
 =&\,\mathrm{sdet}\bigg((1-BC) {1\over 1-ZC}-(1-BC){1\over 1-ZC}Z\tilde Z\bigg)
 =\,\mathrm{Sdet
}\bigg((1-BC) {1\over 1-ZC}(1-Z\tilde Z)\bigg)\,.
\end{align}
We have indeed arrived at the right-hand side of (\ref{eq:79}) and
proven that assertion. Our construction of the color-flavor
transformation is thus completed.

\section{Flat measure  $d(Z,\tilde Z)$}
\label{app:flat}

For the sake of completeness, the flatness of the measure $d(Z,\tilde
Z)$ should be explained. Following standard practice (see,
e.g. \cite{Haake10}) we get that measure form the metric in the superspace
accommodating the matrix $Q=T\Lambda T^{-1}$,
\begin{align}
  \label{eq:81}
  \mathrm{Str}\,dQ^2=\mathrm{Str}\,[T^{-1}dT,\Lambda]^2\,.
\end{align}
For our 'rational parametrization' of the transformation,
$T=\big({0\atop\tilde Z}{Z\atop 0}\big)$, we have
\begin{align}
  \label{eq:63}
  [T^{-1}dT,\Lambda]=\bigg({0\atop 2(1-\tilde ZZ)^{-1}d\tilde Z}
                                         {-2(1-Z\tilde Z)^{-1}dZ\atop 0}\bigg)
\end{align}
 and therefore
\begin{align}
  \label{eq:82}
  \mathrm{Str}\,dQ^2\propto\,\mathrm{str}\,(1-Z\tilde Z)^{-1}dZ
  (1-\tilde Z Z)^{-1}d\tilde Z \,; \end{align} here we have dropped a
numerical factor which can be fixed in the end by normalizing the
measure.  It is now convenient to block diagonalize as
\begin{align}
  \label{eq:83}
  (1-Z\tilde Z)^{-1}=ARA^{-1}\qquad \mathrm{and}\qquad
  (1-\tilde Z Z)^{-1}=DRD^{-1}\,.
\end{align}
We here meet an analogy with the singular-value decomposition
(\ref{eq:singvaldec}) which in the present case makes $A^{-1}ZD$ and
$D^{-1}\tilde ZA$ block diagonal. Not fortuitously, the present  block
diagonalizing transformation is a special case of (\ref{eq:6}) (with
$B=C=0$). We should add that  $A$, $D$, and $R$ are $2\times 2$ in BF
and $R=\big({R_{BB}\atop}{\atop R_{FF}}\big)$ with the $N\times N$
blocks $R_{BB},R_{FF}$ in QD. The 'squared length element' for $dQ$
then becomes
\begin{align}
  \label{eq:84}
  \mathrm{Str}\,dQ^2\propto \mathrm{str}RA^{-1}dZDRD^{-1}d\tilde ZA
                               \equiv  \mathrm{str}RdYRd\tilde Y
\end{align}
or, after writing out the supertrace in BF explicitly,
\begin{align}
  \label{eq:85}
   \mathrm{Str}\,dQ^2\propto \mathrm{tr_{QD}}\Big(
    &R_{BB}dY_{BB}R_{BB}d\tilde Y_{BB} + R_{BB}dY_{BF}R_{FF}d\tilde Y_{FB}
       \\ \nonumber
    &-R_{FF}dY_{FB}R_{BB}d\tilde Y_{BF} - R_{FF}dY_{FF}R_{FF}d\tilde Y_{FF}
    \Big)\,.
\end{align}
Next, we introduce four-component vectors whose elements are $N\times
N$ in QD to write
\begin{align}
  \label{eq:86}
  \left(
  \begin{array}{c}dW_{BB}\\dW_{FF}\\dW_{BF}\\dW_{FB}\end{array}
  \right)\equiv
  \left(
  \begin{array}{cccc}
  R_{BB}&0&0&0\\0&-R_{FF}&0&0\\0&0&R_{BB}&0\\0&0&0&-R_{FF}
  \end{array}
  \right)
  \left(
  \begin{array}{c}dY_{BB}\\dY_{FF}\\dY_{BF}\\dY_{FB}\end{array}
  \right)\,.
\end{align}
Momentarily employing a short hand with $dW$ and $dY$  four-component
supervectors, $\hat R$ a $4\times 4$ supermatrix, and all entries $N\times
N$ in QD, we can write the foregoing transformation from $dY$ to $dW$
as $dW=\hat RdY$. Obviously now, that transformation has the
Berezinian unity since $\mathrm{Sdet}\hat R=1$. Analogously, we get
$d\tilde W=\hat R d\tilde Y$ with unit Berezinian.

Proceeding to the transformations $dY=A^{-1}dZD$ and $d\tilde
Y=D^{-1}\tilde ZA$ we propose to reveal the (nearly obvious) fact that
the respective Berezinians are reciprocal such that the combination of
these two transformation has unit Berezinian. The automatic conclusion
will then be the flatness of the measure $d(Z,\tilde Z)$ since in the
sequence of transformations $dQ\to dWd\tilde W\to dYd\tilde Y\to
dZd\tilde Z$ flatness is preserved step by step.

So we look at $dY=A^{-1}dZD$ and write that transformation as a
sequence of two, $dY=dZ'D$ with $Z'=A^{-1}dZ$. Writing the first step
more explicitly, $dY_{ij}=\sum_k dZ'_{ik}D_{kj}$ where the indices
comprise BF and QD labels we face $2N$ transformations, one for each
'value' of $i$. The overall Berezinian is
$\big(\mathrm{Sdet}D\big)^{2N}$. Arguing similarly we find that the
second step has the Berezinian $\big(\mathrm{Sdet}A^{-1}\big)^{2N}$
such that their combination brings about the product
$\big({\mathrm{Sdet}D\over \mathrm{Sdet}A}
\big)^{2N}$. The transformation $d\tilde
Y=B^{-1}\tilde ZA$ has $A$ and $D$ swapped such that it indeed has the
reciprocal Berezinian. We are done.

\section{Invariances of the $Q$ manifold}
\label{app:Qinvariance}

We had already seen in Sect.~\ref{subsec:fluctuations} that the
product of two good transformations $T=\Big({1\atop \tilde Z}{Z\atop
  1}\Big)$ is again good. We here extend the previous argument with
the goal of establishing the transformation (\ref{eq:6}) as the
general invariance of the manifold $Q=T\Lambda T^{-1}$. We start with
the product
\begin{align}
  \label{eq:87}
  \bigg({A\atop B}{C\atop D}\bigg)\bigg({1\atop \tilde Z}{Z\atop 1}\bigg)
=\bigg({A+B\tilde Z\atop C+D\tilde Z}\;{AZ+B\atop CZ+D}\bigg)
\equiv T'\; \bigg({(A+B\tilde Z)^{-1}\atop }\;{\atop (CZ+D)^{-1}}\bigg)\,.
\end{align}
The block diagonal matrix factored out on the right of $T'$ provides
$T'$ with unit diagonal blocks as required for a good transformation
placing $T'\Lambda T'^{-1}$ into the manifold of $Q$'s. Inasmuch as
the right cofactor of $T'$ commutes with $\Lambda$ we may say that
$T'$ and $ \big({A\atop B}{C\atop D}\big)\,T$ generate the same $Q$.
Moreover, we have
\begin{align}
  \label{eq:88}
  T'=\bigg({1\atop \tilde Z'}{Z'\atop 1}\bigg)
\end{align}
with $Z'$ and $\tilde Z'$ related to $Z$ and $\tilde Z$ by
(\ref{eq:6}).

We hurry to add that the matrices $A,B,C,D$, all $2N\times 2N$ in
BF$\otimes$QD, are not completely arbitrary. As already mentioned they
must allow invertibility of the AR supermatrix $\big({A\atop C}{B\atop
  D}\big)$. They are further restricted such that $Z'$ and $\tilde Z'$
retain the properties ${\tilde Z}'_{BB}={Z'}_{BB}^\dagger$, ${\tilde
  Z}'_{FF}=-{Z'}_{FF}^\dagger$, and $|{Z'}_{BB}{Z'}_{BB}^\dagger|<1$.

\section{$Q$ manifold as a symmetric space}
\label{app:symspace}

We here sketch how group theory provides a natural setting for
everything said in the preceeding three appendices.
The group $GL(2N/2N)$ consists of invertible $4N \times 4N$
supermatrices (over complex numbers) which  we write as
\begin{equation}
  g= \begin{pmatrix}
    A & B\\
    C & D
  \end{pmatrix}\,;
\end{equation}
each block is a $2N \times 2N$ supermatrix which can in turn be
written with $N \times N$ blocks, e.g.  $A=\big( { A_{BB}\atop
  A_{FB}}{ A_{BF} \atop A_{FF} }\big)$. Here $A_{BB}$ and $A_{FF}$
contain commuting numbers (complex numbers and even elements of a
Grassmann algebra with complex coefficients), while $A_{BF}$ and
$A_{FB}$ contain anti-commuting numbers (odd elements of a Grassmann
algebra with complex coefficients). Otherwise the blocks $A$, $B$, $C$
and $D$ are only restricted by invertibility of the matrix $g$.

We define the set of supermatrices of the form
\begin{equation}
  Q(g)= g \Lambda g^{-1}\qquad \mathrm{with}\qquad 
\Lambda= \bigg({1 \atop 0}{ 0 \atop -1} \bigg)\,,
\end{equation}
the coset space $\mathcal{M}_N=GL(2N/2N)/ GL(N/N)^2$. Indeed, the
matrix $\Lambda$ commutes with all block diagonal $4N\times 4N$
supermatrices whose diagonal blocks are $2N\times 2N$ and belong to
the subgroup $GL(N/N)$. Moreover, the manifold in question is a
$GL(2N/2N)$-invariant space. These properties reveal
$\mathcal{M}_N$ as what is called a symmetric space.
Introducing a 'coordinate chart' which parametrizes almost all
matrices in $\mathcal{M}_N$ we write
\begin{equation}
  Q(Z,\tilde{Z})= T(Z,\tilde{Z}) \Lambda T(Z,\tilde{Z})^{-1}
\qquad \mathrm{and}\qquad
  T=\begin{pmatrix}
    1 & Z\\
    \tilde{Z} & 1
       \end{pmatrix}
\end{equation}
where $Z$ and $\tilde{Z}$ are independent $2N \times 2N$-supermatrices
(over complex numbers).

For further clarification we look at the numeric (non-nilpotent) part
of $Q(Z,\tilde{Z})$ by setting all Grassmannian generators to zero.
This implies $Z_{FB}=Z_{BF}=\tilde{Z}_{BF}=\tilde{Z}_{FB}=0$, while
the remaining entries become purely numeric (complex).  The ensuing
matrix $T(Z,\tilde{Z})^{(\mathrm{num})}$ contains two uncoupled
subblocks. We thus see that the manifold of matrices
$Q(g)^{(\mathrm{num})}$ where $g$ runs over $GL(2N/2N)$ is equivalent
to $GL(2N,\mathbb{C})^2/GL(N,\mathbb{C})^4$.  In the color-flavor
transformation an integral over the symmetric space $\mathcal{M}_N$ is
used. The integration requires a choice of the integration range for
the numeric part $Q(g)^{(\mathrm{num})}$. In our case (one color and
$N$ flavors) the right choice is the product of two symmetric spaces,
the non-compact $U(N,N)/U(N)^2$ for the Bose sector and the compact
$U(2N)/U(N)^2$ for the Fermi sector.  The non-compact part is spanned
by $Z_{BB}$ and $\tilde{Z}_{BB}$ with restrictions
$Z_{BB}^\dagger=\tilde{Z}_{BB}$ and all eigenvalues of $Z_{BB}^\dagger
Z_{BB}$ strictly inside the unit circle. The compact part is spanned
by $Z_{FF}$ and $\tilde{Z}_{FF}$ with the restriction
$Z_{FF}^\dagger=-\tilde{Z}_{FF}$.

Generalizing the standard Riemann geometry of symmetric spaces spanned
by ordinary matrices to supermatrices \cite{Zirnb96} we find the flat
integration measure as shown above in Appendix \ref{app:flat}.

\section{Existence worries}
\label{app:existence}

Zirnbauer has argued \cite{Zirnb99} that the sigma model has $N$ 'zero
modes', the projectors onto the eigenstates of $U$, forbidding
preponderance of supermatrices $Z,\tilde Z$ with smooth Wigner
functions in the semiclassical limit. Even though there is one smooth
stationary Wigner function, the one corresponding to the classical
ergodic equilibrium, the $N-1$ others do have quantum fine
structure. Moreover, according to Zirnbauer, the latter states make
for 'neutral directions' at $Z=\tilde Z=0$ allowing for huge
fluctuations.

To appreciate the problem we momentarily employ the
eigenrepresentation of $U$ for the supermatrices $Z,\tilde Z$ in the
sigma model integral (\ref{eq:10},\ref{eq:11}). In that representation
the potentially dangerous zero modes can be scrutinized most
easily. If we could do the superintegral exactly we would expect to
get the formally exact expression (\ref{eq:C(e)}) for the correlator
in terms of the eigenvalues of $U$, no more and no less. While that
expression becomes a distribution for real $e$, an imaginary part
${1\over N}\ll\mathrm{Im}\,e\ll1$ turns the correlator into a smooth
self-averaging function \cite{Braun14}. So protected, the
superintegral for $\mathcal Z$ should and will in fact turn out to be
well behaved.

We begin our work with the quadratic part of the action, thus picking
up fluctuations about the \textit{standard saddle}
$Z=\tilde Z=0$ at which, acording to (\ref{eq:susySwithQ}),
$Q=\Lambda$ and $\mathcal S=0$. Abbreviating the eigenphase spacings
as $\Delta_{\mu\nu}=\phi_\mu-\phi_\nu$ and accounting for $\tilde
Z_{BB}= Z_{BB}^\dagger,\,\tilde Z_{FF}=-Z_{FF}^\dagger$ we get
\begin{align}
  \label{eq:Sss}
  \mathcal S^{(\mathrm{stand})}=&\;\mathrm{str}\Big(\tilde Z Z-U\tilde ZU^\dagger
  \textstyle\big({b\atop}{\atop d}\big)Z\big(\textstyle{a\atop}{\atop c}\big)\Big)\\
 \nonumber =&\sum_{\mu,\nu=1}^N \Big(
   |Z_{BB\nu\mu}|^2(1-ab\,\e^{-\I\Delta_{\mu\nu}})
 +|Z_{FF\nu\mu}|^2(1-cd\,\e^{-\I\Delta_{\mu\nu}})\\ \nonumber
 &\quad\qquad+\tilde Z_{BF\mu\nu}Z_{FB\nu\mu}(1-ad\e^{-\I\Delta_{\mu\nu}})
  -\tilde Z_{FB\mu\nu}Z_{BF\nu\mu}(1-bc\e^{-\I\Delta_{\mu\nu}})
\Big)\,.
\end{align}
The pertinent Gaussian integral $\mathcal Z^{(\mathrm{stand})}=\!\int\!\!
d(Z,\tilde Z)\,\e^{-S^{(\mathrm{stand})}}$ gives
\begin{align}
  \label{eq:Zss}
   \mathcal Z^{(\mathrm{stand})}=
  \prod_{\mu,\nu=1}^N\;{\big(1-ad\,\e^{-\I\Delta_{\mu\nu}}\big)
                               \big(1-bc\,\e^{-\I\Delta_{\mu\nu}}\big)
   \over \big(1-ab\,\e^{-\I\Delta_{\mu\nu}}\big)
   \big(1-cd\,\e^{-\I\Delta_{\mu\nu}}\big)}\;=\,
  \prod_{\mu,\nu=1}^N\left(1+{(a-c)(b-d) \e^{-\I\Delta_{\mu\nu}}
   \over \big(1-ab\,\e^{-\I\Delta_{\mu\nu}}\big)
   \big(1-cd\,\e^{-\I\Delta_{\mu\nu}}\big)}\right).
\end{align}
The general structure $\mathcal Z-1\propto (a-c)(b-d) $ is obviously
respected. More importantly, we have $cd=ab{c\over a}{d\over b}$ with
$|{c\over a}|=|{d\over b}|=1,\,ab=\e^{\I 2e/N}$ and can conclude that
the integral exists as long as Im\,$e>0$.

To simplify the study of the limit Im\,$e\downarrow 0$ we
extract the Gaussian approximation to the correlator,
\begin{align}
  \label{eq:Css}
  C^{(\mathrm{stand})}(e)={2a^2\over N^2}{\mathcal Z^{(2)}-1\over
                 (a-b)(c-d)}\Big|_{a=b=c=d=\e^{\I e/N}}\;
    =\,{2\over N^2}\sum_{\mu,\nu=1}^N{\e^{\I(2e/N-\Delta_{\mu\nu})}
                         \over \big(1-\e^{\I(2e/N-\Delta_{\mu\nu})}\big)^2}\,.
\end{align}
Instead of producing the train of Dirac deltas characteristic for the
real part of the exact correlator, the limit $\epsilon\downarrow 0$
here gives derivatives of Dirac deltas. Only the third
primitive is finite and continuous. The zero modes
($\mu=\nu$) are no more 'unruly' than the others. For  Im$e\sim{1\over N}$
the contribution of the standard saddle is well behaved.

Other saddles than the standard $\Lambda$ can contribute. Varying the
action (\ref{eq:susySwithQ}) w.r.t. $Z$ and $\tilde Z$ one finds
saddle-point equations with many more solutions obtained by permuting
elements of $\Lambda$ such that a number of entries $+1$ are shifted
from the retarded block to the advanced one and the same number of
entries $-1$ from the advanced to the retarded block, the shifts possibly
in different 'eigensectors' $\mu$. One is thus led to the so
called \textit{Andreev-Altshuler saddles}
\cite{Andre95}. Of lesser interest for us are saddles contributing
finitely to $\mathcal Z$ but not to the correlator $C(e)$ -- due to
the appearance of higher powers of $(a-c)$ and/or $(b-d)$. We focus on
the particlar ones relevant for $C(e)$, and these differ from the
standard $\Lambda$ by just a single swap of a $+1$ and a $-1$ between
the retarded and the advanced block, either between two different
eigensectors $\mu$ and $\mu'$ or within a single eigensector $\mu$. We
shall see that the swaps in question happen only within the Fermi
sector such that the pertinent permutations are
\begin{align}
  \label{eq:perm}
  R=P_B+P_F\otimes\big((1-P)+P\otimes\tau_1\big)\,\qquad \mathrm{with}
  \qquad R^2=1\,;
\end{align}
here $P_B$ and $P_F$ respectively project onto the Bose and Fermi
sector while $P$ projects onto the subspace with fixed $\mu$ and
$\mu'$, either coinciding ($\mu=\mu'$) or different.  We refrain from
labeling $R$ and $P$ with these indices.  In the end, we shall sum
over the contributions of all $N^2$ such pairs. In the Fermi sector of
the ($\mu\mu'$)-subspace a two dimensional AR space is left and
therein the swap we are after is realized by the Pauli matrix
$\tau_1=\big({\atop 1}{1\atop}\big)$.  Each such permutation $R$ turns
the standard saddle $\Lambda$ into its $R\Lambda R$.

Fluctuations about these non-standard saddles are conveniently
parametrized by 'permuting from the outside'. Working with the form
(\ref{eq:susySwithQ}) of the action we parametrize as
$Q^{(R)}=RQR=RT\Lambda T^{-1}R$ with $T$ as before -- rather than with 
$T^{(R)}R\Lambda R(T^{(R)})^{-1}$ and a new transformation $T^{(R)}$.
Upon substituting the new saddle $RQR$ into the action
(\ref{eq:susySwithQ}) we obtain
\begin{align}\label{eq:SR}
  \mathcal S(Z,\tilde Z)=\mathrm{Str}\ln\Big(1+ \hat X^{(R)} Q(Z,\tilde Z)\Big)
                      +\mathrm{Str}\ln(1+\hat U), 
  \qquad 
  \hat X^{(R)}\equiv R \hat X R^{-1},\qquad \hat X = \Lambda (1-\hat U)(1+\hat U)^{-1}\,.
\end{align}
Inasmuch as we are working with the eigenrepresentation of the Floquet
matrix $U$, the matrix $\hat X^{(R)}$ is diagonal in
AR$\otimes$BF$\otimes$QD. Indeed, writing $R=1+PP_F(-1+\tau_1)$ we
have $\hat X^{(R)}=\hat X+PP_F(-\hat X+\tau_1X\tau_1)$; now since
$\hat X=\big({X_+\atop}{\atop X_-}\big)$ is diagonal, so is $\hat X^{(R)}=
\big({X_+\atop}{\atop X_-}\big)+PP_F \big({X_--X_+\atop}{\atop X_+- X_-}\big)$.
To capture fluctuations we expand the action (\ref{eq:SR}) to the
order $Z\tilde Z$. A straightforward calculation yields
\begin{align}
  \label{eq:20}
  \mathcal S^{(R)}=\mathcal S^{(R)}_0
        +\mathrm{str}\big(Z\tilde Z-\hat Y^{(R)}_+Z \hat Y^{(R)}_-\tilde Z\big)
\end{align}
where $\mathcal S^{(R)}_0=\mathrm{Str}\ln\big(1+\hat
X^{(R)}\Lambda\big) +\mathrm{Str}\ln\big(1+\hat U\big)$ is the action
at the saddle and $\hat Y^{(R)}=2\big((\hat
X^{(R)})^{-1}\Lambda+1\big)^{-1}-1$ a matrix inheriting diagonality
from $\hat X^{(R)}$; the retarded/advanced blocks $\hat Y^{(R)}_\pm$
of that matrix appear in the foregoing 'quadratic' action. Evaluating
the quantity $\mathcal S^{(R)}_0$ is an easy matter, again due to the
diagonality of $\hat U$ and $\hat X^{(R)}$,
\begin{align}
  \label{eq:42}
  \mathcal S^{(R)}_0=-\ln\big(cd\e^{-\I\Delta_{\mu\mu'}}\big)\,.
\end{align}

We now integrate over $Z,\tilde Z$ as for the standard saddle and get
\begin{align}
  \label{eq:ZR}
  \mathcal Z^{(R)}=\big(cd \e^{-\I\Delta_{\mu\mu'}}\big)\,\prod_{\rho,\sigma=1}^N
  \,{\big(\hat Y^{(R)}_{+B\rho}\hat Y^{(R)}_{-F\sigma}-1\big) 
   \big(\hat Y^{(R)}_{+F\rho}\hat Y^{(R)}_{-B\sigma}-1\big)\over 
   \big(\hat Y^{(R)}_{+B\rho}\hat Y^{(R)}_{-B\sigma}-1\big) 
   \big(\hat Y^{(R)}_{+F\rho}\hat Y^{(R)}_{-F\sigma}-1\big)}
   \,\equiv\, \big(cd \e^{-\I\Delta_{\mu\mu'}}\big)\,\prod_{\rho,\sigma=1}^N\Xi_{\rho\sigma}\,.
\end{align}
This result formally contains the standard saddle, as one can check by
letting $R$ be the identity whereupon one comes back to $\mathcal
S_0=0$ and $\hat Y=2\big((\hat X^{-1}\Lambda+1\big)^{-1}-1$; the
entries in the diagonal matrix $\hat Y$ are easily found and
(\ref{eq:Zss}) is recovered. But now we are interested in the
non-standard saddles generated by the permutation (\ref{eq:perm}) for
which only two of the diagonal elements of $\hat Y^{(R)}$ differ from
the ones of $\hat Y$,
\begin{align}
  \label{eq:40}
  \hat Y^{(R)}_{+F\mu}=-\textstyle{1\over d}\,\e^{-\I\phi_{\mu'}}
  \quad\qquad \mathrm{and}\qquad\quad
  \hat Y^{(R)}_{-F\mu'}=-\textstyle{1\over c}\,\e^{\I\phi_{\mu}}\,;
\end{align}
relative to the elements of $\hat Y$ the change is $c\to{1\over d}$,\,
 $d\to{1\over c}$, and $\mu\leftrightarrow \mu'$. The factor $\Xi_{\mu\mu'}$ 
in (\ref{eq:ZR}) thus becomes
\begin{align}
  \label{eq:41}
  \Xi_{\mu\mu'}=-{(a-c)(b-d) \e^{-\I\Delta_{\mu\mu'}}
   \over \big(1-ab\,\e^{-\I\Delta_{\mu\mu'}}\big)
   \big(1-cd\,\e^{-\I\Delta_{\mu\mu'}}\big)}\,,
\end{align}
indeed proportional to $(a-c)(b-d)$. We can therefore set $a=c$ and
$b=d$ in all other factors $\Xi_{\rho\sigma}$ in (\ref{eq:ZR}) which
thereby all take the value 1. Summing over $\mu$ and $\mu'$ in
(\ref{eq:ZR}) to pick up the contributions of all these saddles we get 
\begin{align}
  \label{eq:43}
  \mathcal Z^{(AA)}=-\sum_{\mu\mu'}{(a-c)(b-d) \e^{-2\I\Delta_{\mu\mu'}}
   \over \big(1-ab\,\e^{-\I\Delta_{\mu\mu'}}\big)
   \big(1-cd\,\e^{-\I\Delta_{\mu\mu'}}\big)}+\ldots\,,
\end{align}
where the dots refer to contributions $\propto (a-c)^m(b-d)^n$ with
$m+n>1$; the latter are uncapable of influencing the correlator for
which reason we do not bother to treat the pertinent further saddles.

For the correlator we finally get
\begin{align}
  \label{eq:44}
  C(e)={2a^2\over N^2}{\mathcal Z^{(\mathrm{stand})}+ \mathcal Z^{(AA)}-1
          \over (a-c)(b-d)}\Big|_{a=b=c=d=\e^{\I e/N}}
        ={2\over N^2}\sum_{\mu\mu'}{\e^{\I(2e/N-\Delta_{\mu\mu'})}
           \over 1-\e^{\I(2e/N-\Delta_{\mu\mu'})}}\,,
\end{align}
the exact expression (\ref{eq:C(e)}) in terms of the
eigenrepresentation of $U$. The formal exactness of the saddle-point
treatment is somewhat of a surprise and may be worth further thought.

Most importantly in our present context, one need not worry about
existence of the sigma model. As long as Im\,$e>0$ there are no
massless modes invalidating expansions to higher orders in $Z,\tilde
Z$. Smoothing by Im\,$e>0$ is necessary anyway, to turn the correlator
into a smooth function. Moreover, in the train of Dirac deltas arising
for Im$e\downarrow 0$ the formally massless diagonal modes
($\mu=\mu'$) are in no way distinguished over the formally massive
off-diagonal ones. Unfortunately, the $U$-representation is useless
inasmuch as it yields no clues as to why and under what conditions the
smoothed correlator behaves universally.  The foregonig investigation
in fact lends support to endeavors involving representations which in
contrast to the $U$-representation invite implementation of the
semiclassical limit.

\section{Derivation of Eq.~\protect{\ref{eq:susySwithQ}}}
\label{sec:susySwithQ}

The action in (\ref{eq:11}) can be rewritten as
\begin{eqnarray}
  \label{eq:33a}
S=-\,\mathrm{Str}\ln\left(1-\bigg({0 \atop Z}
{\tilde Z \atop 0}\bigg)\right)+\,\mathrm{Str}\ln\left(1-
\bigg({ U \hat e_+\atop 0} {0 \atop U^\dagger \hat e_-}\bigg)
\bigg({0 \atop  Z} {\tilde Z \atop 0}\bigg)\right)
\end{eqnarray}
due to $\mathrm{Str}\left({0 \atop \tilde Z} {Z \atop
    0}\right)^{2n+1}=0$ and the Taylor expansion $\ln (1-x)=-\sum
{x^n\over n}$.
We now use the freedom to rename as $Z\leftrightarrow \tilde Z$ and
then employ the identity
\begin{equation}
  \label{eq:34}
  1-\left({0 \atop \tilde Z}{Z \atop 0}\right)=2\left(1+Q\Lambda\right)^{-1}\,,
  \quad \Lambda =\left({1 \atop 0}{0 \atop -1}\right)\,,
\end{equation}
which is easily verified upon inserting the definition of $Q$ given in
(\ref{eq:susySwithQ}) and  the matrix inverse
\begin{equation}
  \label{eq:35}
  \left({1 \atop \tilde Z}{Z \atop 1}\right)^{-1}=
  \left({1 \atop -\tilde Z}{-Z \atop 1}\right)
  \left({\frac{1}{1-Z\tilde Z} \atop 0}{0 \atop \frac{1}{1-\tilde Z Z}}\right)\,.
\end{equation}
The action thus becomes
\begin{eqnarray}
  \label{eq:36}
  S=-\,\mathrm{Str}\ln\frac{2}{Q\Lambda+1}
          +\,\mathrm{Str}\ln\left(1-
      \Big({ U \hat e_+\atop 0} {0 \atop U^\dagger \hat e_-}\Big)
      \frac{Q\Lambda-1}{Q\Lambda+1}\right)
    =\mathrm{Str}\ln\left((1-\hat U) Q\Lambda
          +(1+\hat U)\right)\,.
\end{eqnarray}
Employing the shorthand $\hat U$ introduced in (\ref{eq:Q}) and
extracting the factor $1+\hat U$ from the argument of the logarithm we
arrive at $ S(Z,\tilde Z)=\mathrm{Str}\ln \left[1+(1-\hat U)(1+\hat U)^{-1}Q \Lambda\right]
\,+ \;\mathrm{Str}\ln (1+\hat U)$, as given in (\ref{eq:susySwithQ}).

\section{Evaluation of the Grassmann integral (\ref{eq:G})}
\label{app:intZ0}

An economic way of doing the Grassmann integral $\mathcal G$ is by
differentiation, $\int d\eta^*d\eta d\tau^* d\tau
(\cdot)=\partial_{\eta^*}\partial_{\eta}\partial_{\tau^*}\partial_\tau(\cdot)$,
since the chain and product rules prove helpful.  It is worthwhile to
spell out the product rule since a peculiarity of Grassmann calculus
must be pointed at. Due to the anticommutativity of Grassmannians we
must distinguish between ''even'' and ''odd'' functions
$A(\tau,\tau^*,\eta,\eta^*)$.  Even ones additively contain only
bilinear terms and/or the quadruple of the arguments while odd
functions contain only linear or trilinear terms. The product rule can
then be written as
\begin{align}
  \label{eq:67}
  \partial_\tau \Big[A(\tau,\tau^*,\eta,\eta^*)B(\tau,\tau^*,\eta,\eta^*)\Big]=
    A_\tau(\tau^*,\eta,\eta^*)B(\tau^*,\eta,\eta^*)
  \pm A(\tau^*,\eta,\eta^*) B_\tau(\tau^*,\eta,\eta^*)
\end{align}
where the plus (minus) sign refers to even (odd) $A$. The absence of
the argument $\tau$ on the right-hand side of the forgoing rule and
the alternative $\pm$ constitute the announced peculiarity. We promise
to pedantically adhere to the notation thus
introduced, while evaluating $\mathcal G $: the
absence of any of the four arguments from any function means either
that the argument has been removed by differentiation
(as indicated by the corresponding index) or that a cofactor is
differentiated (whereupon in the function in question the argument
must be replaced with $0$ as well).

Writing
$D(\tau,\tau^*,\eta,\eta^*)=\mathrm{Sdet}\,\big(1-m(\tau,\tau^*,\eta,\eta^*)\big)$
we have $\mathcal
G=\partial_{\eta^*} \partial_{\eta}\partial_{\tau^*}\partial_{\tau}D(\tau,\tau^*,\eta,\eta^*)^{-N}$.
In differentiating we respect that the superdeteminant
$D(\tau,\tau^*,\eta,\eta^*)$ is even, to get
\begin{align}
  \label{eq:68}
  \mathcal G=&\,\partial_{\eta^*} \partial_{\eta}\partial_{\tau^*}
\big[-N D_\tau(\tau^*,\eta,\eta^*)D(\tau^*,\eta,\eta^*)^{-(N+1)}\big]
\\ \nonumber
=&\,\partial_{\eta^*} \partial_{\eta}
\big[-N D_{\tau^*\tau}(\eta,\eta^*)D(\eta,\eta^*)^{-(N+1)}
       -N(N+1)D_\tau(\eta,\eta^*)D_{\tau^*}(\eta,\eta^*) D(\eta,\eta^*)^{-(N+2)}\big]
\,;
\end{align}
in the last step we have used that $D_\tau(\tau^*,\eta,\eta^*)$ is
odd. Continuing in this vein we get a sum of 15 terms, all with
the four derivatives  distributed over up to four
superdeterminants. However, eleven of these terms vanish due to
$D_\tau=D_{\tau^*}=D_\eta=D_{\eta^*}=0$. Indeed, a glance at the
definitions of the matrices $\Delta_\pm$ in (\ref{eq:61}) reveals that
upon setting three of the four Grassmannians to zero, one of these
matrices must vanish while the other becomes purely off-diagonal, such
that the supertrace $\mathrm{Str}\ln(1-m)$ vanishes; moreover, that
supertrace is devoid of the bilinear summands $\tau\eta$ and
$\tau^*\eta^*$ and we conclude $D_{\eta\tau}=D_{\eta^*\tau^*}=0$. We
finally use $D=D(0,0,0,0)=1$ and are left with
\begin{align}
  \label{eq:70}
  \mathcal G=-ND_{\eta^*\eta\tau^*\tau}
                    +N(N+1) \big(D_{\tau^*\tau}D_{\eta^*\eta}
                    +D_{\eta^*\tau}D_{\eta\tau^*}\big)\,.
\end{align}
We must now inspect the $2\times 2$ superdeterminant
$D(\tau,\tau^*,\eta,\eta^*)$. Momentarily dropping
 arguments we write
\begin{align}
  \label{eq:69}
  D=\mathrm{Sdet}\left({1-m_{BB}\atop -m_{FB}}{-m_{BF}\atop 1-m_{FF}}\right)
=\frac{1-m_{BB}}{1-m_{FF}} -\frac{m_{BF}m_{FB}}{(1-m_{FF})^2}\,.
\end{align}
By its definition (\ref{eq:m}), the matrix $m$ has even
diagonal entries and odd off-diagonal ones. Furthermore, all four
entries are nilpotent. In particular, $m_{FF}^n=0$ for $n>2$,
and therefore we can expand as
\begin{align}
  \label{eq:72}
  D-1&=-m_{BB}+m_{FF}-m_{BB}m_{FF}+m_{FF}^2-m_{BF}m_{FB}(1+2 m_{FF})\,.
\end{align}
That expansion must equal
$D(\tau,\tau^*,\eta,\eta^*)=1+D_{\tau^*\tau}\tau\tau^*+D_{\eta^*\eta}\eta\eta^*
+D_{\eta^*\tau}\tau\eta^*+D_{\eta\tau^*}\tau^*\eta
+D_{\eta^*\eta\tau^*\tau}\tau\tau^*\eta\eta^*$, with the coefficients
appearing in (\ref{eq:70}); note that we are back to the promised
pedantery. To determine these coefficients we need to inspect the
matrix $m$ in detail. Momentarily writing
$\delta_+=a-c,$ and $\delta_-=b-d$ we have
\begin{align}
  \label{eq:mab}
  m_{BB}(\tau,\tau^*,\eta,\eta^*)&=(1-abl_B)^{-1}\Big(l_B\big(-\delta_+b\tau\tau^*
               +\delta_-a\eta\eta^*-\delta_+\delta_-\tau\tau^*\eta\eta^*\big)
               -\delta_+\delta_-\sqrt{-l_Bl_F}\,\tau^*\eta\Big)\\ \nonumber
  m_{FF}(\tau,\tau^*,\eta,\eta^*)&=(1-cdl_F)^{-1}\Big(l_F\big(-\delta_+d\tau\tau^*
               +\delta_-c\eta\eta^*-\delta_+\delta_-\tau\tau^*\eta\eta^*\big)
               -\delta_+\delta_-\sqrt{-l_Bl_F}\,\tau \eta^*\Big)\\ \nonumber
   m_{BF}(\tau,\tau^*,\eta,\eta^*)&=
                (1-abl_B)^{-1}\Big(l_B\delta_+\big(b+\delta_-\eta\eta^*\big)\tau
               +\sqrt{-l_Bl_F}\,\delta_-\big(c-\delta_+\tau\tau^*\big)\eta\Big)\\
                 \nonumber
   m_{FB}(\tau,\tau^*,\eta,\eta^*)&=
                (1-cdl_F)^{-1}\Big(l_F\delta_+\big(d+\delta_-\eta\eta^*\big)\tau^*
               +\sqrt{-l_Bl_F}\,\delta_-\big(a-\delta_+\tau\tau^*\big)\eta^*\Big)
\,.
\end{align}
Only the diagonal elements and the product $m_{BF}m_{FB}$ of the
off-diagonals contain terms bilinear in the Grassmannians.
We read out the second derivatives
\begin{align}
  \label{eq:73}
  D_{\tau^*\tau}
                      =\frac{(a-c)(bl_B-dl_F)}{(1-abl_B)(1-cdl_F)}\,,\quad
  D_{\eta^*\eta}
                     =-\frac{(b-d)(al_B-cl_F)}{(1-abl_B)(1-cdl_F)}\,,\quad
  D_{\eta^*\tau}=-D_{\eta\tau^*}
                      =-\frac{(a-c)(b-d) \sqrt{-l_Bl_F}}{(1-abl_B)(1-cdl_F)}
\end{align}
and their combination
\begin{align}
  \label{eq:71}
  D_{\tau^*\tau}D_{\eta^*\eta}+ D_{\eta^*\tau}D_{\eta\tau^*}
&=-\frac{(a-c)(b-d)(abl_B-cdl_F)(l_B-l_F)}{(1-abl_B)^2(1-cdl_F)^2}\,.
\end{align}
The forth derivative, in turn, receives contributions from all terms
on the right-hand side of the expansion (\ref{eq:72}),
\begin{align}
\label{eq:foul}
 D_{\eta^*\eta\tau^*\tau}&=\frac{(a-c)(b-d)(1+cdl_F)(l_B-l_F)}{(1-abl_B)(1-cdl_F)^2}\,.
\end{align}
Putting together Eqs.~(\ref{eq:71}) and (\ref{eq:foul}) we get the announced
result (\ref{eq:G}) for the Grassmann integral $\mathcal G$.

\section{Checking  $\partial_c\partial_d \langle \mathcal S_{\rm
  c}\rangle\big|_{\epsilon_{\pm}=0}\propto{1\over N^2}$ }
\label{app:lastcorr}

We must invoke the easily proven
contraction rule
\begin{align}
    \label{eq:contraction}
\big\langle{\rm str_{BF}} X \tilde z_\mu Y z_\mu\big\rangle_{\rm d}
=&\, \,X_{BB}\left( {Y_{BB}\over 1-ab\e^{-\lambda_\mu}}
                   -{Y_{FF}\over 1-ad\e^{-\lambda_\mu}} \right)
    -X_{FF}\left( {Y_{BB}\over 1-bc\e^{-\lambda_\mu}}
                   -{Y_{FF}\over 1-cd\e^{-\lambda_\mu}} \right)
\\
   =& \,\,{\mathrm {str_{BF}} X\, \mathrm {str_{BF}} Y\over 1-ab\e^{-\lambda_\mu} }
      +{(a-c)b\e^{-\lambda_\mu}\over(1-ab\e^{-\lambda_\mu})^2}X_{FF}\,{\rm str_{BF}}Y
      +{(b-d)a\e^{-\lambda_\mu}\over(1-ab\e^{-\lambda_\mu})^2}Y_{FF}\,{\rm
        str_{BF}}X \label{eq:contraction2} \\ & \nonumber
      +(a-c)(b-d)\,{\e^{-\lambda_\mu}\,(1+ab\e^{-\lambda_\mu})\over
        (1-ab\e^{-\lambda_\mu})^3}X_{FF}Y_{FF}
       +\ldots
\end{align}
with $2\times 2$ matrices $X,Y$ in BF; the dots refer to terms of
higher order in $a-c$ and $b-d$.

Easy to evaluate is the average $\langle \mathrm{str}\,\hat e_+\tilde
z_\mu \hat e_- z_\mu\rangle=\mathcal Z_0^{-1} \int d(B,\tilde
B)\e^{-N\mathcal S_0} \langle\mathrm{str}\hat e_+\tilde z_\mu \hat e_-
z_\mu\rangle_{\rm d}$. The mean-field average herein trivially gives
the factor unity since the quantity to be averaged is independent of
$B,\tilde B$. The contraction rule gives the partial average over the
decaying modes. Due to $\mathrm{str}\,\hat e_+=a-c=\delta_+$ and
$\mathrm{str}\,\hat e_-=b-d=\delta_-$ we immediately get an overall
factor $\delta_+\delta_-$. In the remaining cofactor we may right away
set $a=b=c=d$ since we are interested in $\partial_c\partial_d\langle
\mathcal S_{\rm c}\rangle\big|_{a=b=c=d}$. In particular, we can set
$\mathcal Z_0^{-1} =1$. A few steps of elementary algebra then lead to
$ \langle \mathrm{str}\,\hat e_+\tilde z_\mu \hat e_- z_\mu\rangle
=\delta_+\delta_-(1+a^2\e^{-\lambda_\mu})
(1-a^2\e^{-\lambda_\mu})^{-3}$, independent of $N$. The pertinent
correction to the correlator therefore is of the order ${1\over N^2}$.

We now turn to the average of the second summand under the
supertrace in (\ref{eq:Scmu}),
\begin{align}
  \label{eq:47}
     \Big\langle\mathrm{str}\,\hat e_+{1\over 1-\tilde B \hat B}(1-\tilde B B)
   \tilde z_\mu {1\over 1-\hat B\tilde B}\,\hat e_-(1-B\tilde B) z_\mu
   \Big\rangle
  \equiv
     \langle\mathrm{str}\,X \tilde z_\mu\,Y z_\mu\rangle
\,.
\end{align}
The mean-field average herein is conveniently done with the help of the
singular-value decomposition (\ref{eq:singvaldec}),
\begin{align}
  \label{eq:39}
  \langle\mathrm{str}X\tilde z_\mu Y z_\mu\rangle=\mathcal Z_0^{-1}
  \int_0^1 \!\!\!\!dl_B\!\!\int_{-\infty}^0 \!\!\!\!dl_F
  {1\over (l_B-l_F)^2}\left({(1-l_B)(1-a^2l_F)\over (1-l_F)(1-a^2l_B)}\right)^N
  \!\!\int \!\!d\eta^* d\eta d\tau^* d\tau\e^{-N\mathrm{str}\ln(1-m)}
  \langle\mathrm{str}X\tilde z_\mu Y z_\mu\rangle_{\rm d}\,.
\end{align}
We know already that the foregoing superintegral is proportional to
$(a-c)(b-d)$.  Moreover, for the correlator no terms of higher order
in $a-c$ and $b-d$ are needed. Therefore, we can set $\mathcal Z_0\to
1$ and $a=b=c=d$, except in the resulting factor $(a-c)(b-d)$.

Being after the large-$N$ limit we confine ourselves to small ${e\over
  N}$ such that $a=1+\I{e\over N}+\ldots$. Then the base under the
exponent $N$ becomes proportional to ${1\over N}$ such that we can use
$(1+{x\over N})^N=\e^x(1-{x^2\over 2N}+\ldots)$. For our purposes, we
can drop the ${1\over N}$ correction and write
$\left({(1-l_B)(1-a^2l_F)\over (1-l_F)(1-a^2l_B)}\right)^N=
\exp \big\{2\I e \big ({l_B\over 1-l_B}-{l_F\over 1-l_F}\big) \big\}$
and thus
\begin{align}
  \label{eq:32}
  \langle\mathrm{str}X\tilde z_\mu Y z_\mu\rangle\sim
  \int_0^1 \!\!\!\!dl_B\!\!\int_{-\infty}^0 \!\!\!\!dl_F
  {\exp \big\{2\I e \big ({l_B\over 1-l_B}-{l_F\over 1-l_F}\big) \big\}\over (l_B-l_F)^2}
  \!\!\int \!\!d\eta^* d\eta d\tau^* d\tau\e^{-N\mathrm{str}\ln(1-m)}
  \langle\mathrm{str}X\tilde z_\mu Y z_\mu\rangle_{\rm d}\,.
\end{align}
To continue our quest for the large-$N$ asymptotics we use $\hat
e_+=a-(a-c)P,\,\hat e_-=b-(b-d)P$ with the BF projector
$P=\left({0\atop 0}{0\atop 1}\right)$ to expand the matrices
$(1-\tilde B\hat B)^{-1}$ and $(1-\hat B\tilde B)^{-1}$ in powers of
$\delta_+=(a-c)$ and $\delta_-=(b-d)$,
\begin{align}
  \label{eq:21}
  X&=\hat e_+{1\over 1-\tilde B \hat B}(1-\tilde B B)
    =X_0+\delta_+X_++\delta_-X_-+\delta_+\delta_-X_{+-}+\ldots\,,
\\ \nonumber
  Y&={1\over 1-\hat B\tilde B}\,\hat e_-(1-B\tilde B)
      =Y_0+\delta_+Y_++\delta_-Y_-+\delta_+\delta_-Y_{+-}+\ldots\,,
\end{align}
where the dots stand for higher-order terms incapable of
contributing to the correlator. The matrices
$X_0,X_\pm,X_{+-}$ and $Y_0,Y_\pm,Y_{+-}$ all have finite vanishing
limits as $N\to \infty$.

We proceed to the average over the decaying modes with the help of the contraction rule
(\ref{eq:contraction2}). The shorthand $f_\mu={1-ab\e^{-\lambda_\mu}}$
allows to write
\begin{align}
  \label{eq:25}
  \langle\mathrm{str}X\tilde z_\mu Y z_\mu\rangle_{\rm d}=&
   f_\mu\mathrm{str}X_0\mathrm{str}Y_0\;\\ \nonumber
   &+\delta_+f_\mu\big[\mathrm{str}X_+\mathrm{str}Y_0
                                   +\mathrm{str}X_0\mathrm{str}Y_+
                          +bf_\mu\e^{-\lambda_\mu}X_{0FF}\mathrm{str}Y_0\big]\;
\\ \nonumber
   &+\delta_-f_\mu\big[\mathrm{str}X_-\mathrm{str}Y_0
                                    +\mathrm{str}X_0\mathrm{str}Y_-
\\ \nonumber
  &+\delta_+\delta_-f_\mu\Big[\mathrm{str}X_0\mathrm{str}Y_{+-}
                                           +\mathrm{str}X_{+-}\mathrm{str}Y_0
                                           +\mathrm{str}X_{+}\mathrm{str}Y_-
                                           +\mathrm{str}X_{-}\mathrm{str}Y_+
                                          \\ &\nonumber \hspace{1.5cm}
    +\e^{-\lambda_\mu}f_\mu\big(
                a(\mathrm{str}X_+)Y_{0FF}+a(\mathrm{str}X_0)Y_{+FF}
              +bX_{-FF}\mathrm{str}Y_0+bX_{0FF}\mathrm{str}Y_-\big)
\\ \nonumber& \hspace{1.5cm}
+\e^{-\lambda_\mu}(1+ab\e^{-\lambda_\mu})f_\mu^2X_{0FF}Y_{0FF}\Big]+\ldots
\,.\;
\end{align}
where we should and henceforth shall set $a=b=c=d=\e^{\I e/N}$ except
in the explicit factors $\delta_\pm$.

For the first term in (\ref{eq:25}), the one of zero order in
$\delta_\pm$, we need $X_0=a{1-\tilde BB\over 1-a^2 \tilde BB}\sim
a(1-(2\I e/N)\tilde B B(1-\tilde BB)^{-1})$
and $\mathrm{str}X_0=\mathrm{str}Y_0\sim -{2\I e\over
  N}{l_B-l_F\over (1-l_B)(1-l_F)}$. The full average of the first term
thus reads
\begin{align}
\label{eq:29}
  \langle f_\mu\mathrm{str}X_0\mathrm{str}Y_0\rangle = -f_\mu\left(2e\over N\right)^2
  \int_0^1dl_B \int_{-\infty}^0dl_F \left({l_B-l_F\over (1-l_B)(1-l_F)}\right)^2
\mathcal G{1\over (l_B-l_F)^2}\exp \Big\{2\I e \Big ({l_B\over 1-l_B}-{l_F\over 1-l_F}\Big) \Big\}\,.
\end{align}
Then to leading order in $N$, the Grassmann integral (\ref{eq:154})
becomes $\mathcal G\sim N^2\delta_+\delta_-({l_B\over 1-l_B}-{l_F\over
  1-l_F})^2$. Changing integration variables as $x={ l_B\over
  1-l_B},\,y={l_F\over 1-l_F}$ we get for the average under study
\begin{align}
  \label{eq:31}
   {\langle f_\mu\mathrm{str}X_0\mathrm{str}Y_0\rangle\over (a-c)(b-d) }\sim
-4e^2f_\mu\int_0^\infty dx\int_{-1}^0 dy(x-y)^2\exp\{2\I e(x-y)\}\,,
\end{align}
i.e. independence of $N$.  We conclude the corresponding
correction to the correlator to be $\propto ({e\over N})^2$.

For the full average of the second term of (\ref{eq:25}), the one
linear in $\delta_+$, we need to set $\delta_+=0$ in the nilpotent
matrix $m$ (such that $m\to \delta_-m_-$) and expand as
$\e^{-N\mathrm{str}\ln(1-\delta_-m_-)}=1-N\delta_-\mathrm{str}\,m_-$;
the summand 1 does not survive the Grassmann integration since the
square bracket in the second term of (\ref{eq:25}) contains no
quadrilinear product $\tau\tau^*\eta\eta^*$, such that under the
Grassmann integral we have
$\e^{-N\mathrm{str}\ln(1-\delta_-m_-)}=-N\delta_-\mathrm{str}\,m_-=-\eta\eta^*N(l_B-l_F)$.
The factor $N$ here appearing is compensated by the ${e\over N}$
appearing in the traces $\mathrm{str}X_0=\mathrm{str}Y_0$ such that
again the average $\langle\delta_-f_\mu[\cdot]\rangle$ under study is
asymptotically independent of $N$ and thus incapable of contributing
more than a ${1\over N^2}$ correction to the correlator. Analogous
reasoming gives the same result for the third term in (\ref{eq:25}).

Finally, the fourth term in (\ref{eq:25}), the one bilinear in $\delta
_+$ and $\delta_-$, cannot produce anything larger either since here
the nilpotent matrix $m$ vanishes for $\delta_\pm\to 0$ such that no
factor $N$ comes from $\e^{-N\mathrm{str}\ln(1-\delta_-m)}$
while the square bracket in the fourth term of (\ref{eq:25}) is
asymptotically independent of $N$; indeed, one easily checks that
the terms $\mathrm{str}X_{+}\mathrm{str}Y_-+\mathrm{str}X_{-}\mathrm{str}Y_+$
do not vanish in the limit $N\to\infty$.

\bibliographystyle{unsrt}
\bibliography{fritz}

\end{document}